\newif\ifblind
\blindfalse

\ifblind
\documentclass[conference,blind,balance=true]{IEEEtran}
\else
\documentclass[conference,final,balance=true]{IEEEtran}
\usepackage{flushend} 
\IEEEtriggeratref{18} 
\fi

\makeatletter
\def\mdseries@tt{m}
\makeatother
\usepackage[frozencache=true]{minted}

\usepackage[font=small, justification=centering]{caption}
\makeatletter
\let\MYcaption\@makecaption
\makeatother
\usepackage[font=small]{subcaption}
\makeatletter
\let\@makecaption\MYcaption
\makeatother

\usepackage{float}
\floatstyle{plaintop}
\restylefloat{table}
\usepackage{microtype}
\usepackage{amsfonts}
\usepackage{array}
\usepackage{algorithm}
\usepackage{algorithmic}
\usepackage{graphicx}
\usepackage{multirow}
\usepackage[bottom]{footmisc} 
\usepackage[all]{nowidow} 

\usepackage{booktabs} 

\usepackage{listings}
\DeclareCaptionSubType{listing} 

\newcolumntype{P}[1]{>{\centering\arraybackslash}p{#1}}

\usepackage[hidelinks]{hyperref} 

\fvset{linenos=false, fontsize=\tiny, frame=lines, numbersep=1pt, xleftmargin=.75em}
\usepackage{mathtools} 

\usepackage{multirow} 
\usepackage{enumitem} 
\usepackage{color}
\usepackage[table]{xcolor}

\usepackage[per-mode=symbol]{siunitx}
\DeclareSIUnit \bit {bit}
\DeclareSIUnit \bits {bits}
\DeclareSIUnit \byte {B}
\DeclareSIUnit \bytes {Bytes}
\DeclareSIUnit \cycle {cycle}
\DeclareSIUnit \cycles {cycles}
\DeclareSIUnit \hz {Hz}
\DeclareSIUnit \op {Op}
\DeclareSIUnit \ops {Ops}
\DeclareSIUnit \operand {operand}
\DeclareSIUnit \operands {operands}
\DeclareSIUnit \transfer {T}
\DeclareSIUnit \cell {cell}

\newcommand{\figureref}[1]{\hyperref[fig:#1]{Fig.~\ref{fig:#1}}}
\newcommand{\tableref}[1]{\hyperref[tab:#1]{Tab.~\ref{tab:#1}}}
\newcommand{\listingref}[1]{\hyperref[lst:#1]{Lst.~\ref{lst:#1}}}
\newcommand{\equref}[1]{\hyperref[eq:#1]{Eq.~\ref{eq:#1}}}
\newcommand{\secref}[1]{\hyperref[sec:#1]{Sec.~\ref{sec:#1}}}

\newcommand{\secsymbol}{{\S}}
\newcommand{\seclink}[1]{\hyperref[sec:#1]{\secsymbol\ref{sec:#1}}}

\usepackage{varwidth}
\makeatletter
\newcommand{\breakcaption}{\@dblarg\emit@breakcaption}
\long\def\emit@breakcaption[#1]#2{%
  \expandafter\caption\expandafter[\expandafter\emit@removeafter#1\\\@nil]{%
    \begin{varwidth}[t]{\columnwidth-\widthof{\figurename\space\thefigure:\space}}
    #2
    \end{varwidth}%
  }%
}
\def\emit@removeafter#1\\#2\@nil{#1}
\makeatother

\newminted{c}{}
\newif\ifFirstMintedPart
\makeatletter
\newenvironment{mintedBlock}{%
    \par
    \medskip
    \begingroup
        \setlength{\parskip}{0pt}%
        \setlength{\baselineskip}{0pt}%
        \setlength{\lineskip}{0pt}%
        \let\originalVspace=\vspace
        \renewcommand{\vspace}{\@ifnextchar*\@gobbletwo\@gobble}%
        \setlength{\fboxsep}{0pt}%
        \FirstMintedParttrue
        \noindent\FancyVerbRuleColor{\vrule \@width\linewidth \@height\FV@FrameRule}%
        \originalVspace{2pt}
        \par
}{%
        \par
        \originalVspace{2pt}
        \noindent\FancyVerbRuleColor{\vrule \@width\linewidth \@height\FV@FrameRule}%
        \par
    \endgroup
    \medskip
}
\makeatother
\newenvironment{cpart}[1]{%
    \VerbatimEnvironment
    \ifFirstMintedPart
        \newcommand\currentLineNumber{firstnumber=1}%
    \else
        \newcommand\currentLineNumber{firstnumber=last}%
    \fi
    \newcommand{\beginCCode}{\begin{ccode*}}%
    \expandafter\beginCCode\expandafter{%
        \currentLineNumber,
        frame=none,
        bgcolor=white,
        breaklines, 
        #1
    }%
}{%
    \end{ccode*}%
    \global\FirstMintedPartfalse
}

\usepackage{inconsolata}

\begin{document}

\title{StencilFlow: Mapping Large Stencil Programs\\to Distributed Spatial Computing Systems}

\author{%
\ifblind
\IEEEauthorblockN{Anonymous authors.}
\else
\author{\IEEEauthorblockN{\resizebox{\textwidth}{!}{Johannes de Fine Licht$^*$, Andreas Kuster$^*$, Tiziano De Matteis$^*$, Tal Ben-Nun$^*$, Dominic Hofer$^{*\dagger}$, Torsten Hoefler$^*$}}
\IEEEauthorblockA{$^*$Department of Computer Science, ETH Zurich, Switzerland
$^\dagger$MeteoSwiss, Switzerland\\
\{definelj, kustera, tdematt, tbennun, dohofer, htor\}@ethz.ch}}
\fi
}


\IEEEtitleabstractindextext{%
\begin{abstract}
Spatial computing devices have been shown to significantly accelerate stencil
computations, but have so far relied on unrolling the iterative dimension of a
single stencil operation to increase temporal locality. This work considers the
general case of mapping directed acyclic graphs of heterogeneous stencil
computations to spatial computing systems, assuming large input programs without
an iterative component. StencilFlow maximizes temporal locality and ensures
deadlock freedom in this setting, providing end-to-end analysis and mapping from
a high-level program description to distributed hardware. We evaluate our
generated architectures on a Stratix~10 FPGA testbed, yielding
$\SI{1.31}{}$~$\si{\tera\op\per\second}$ and
$\SI{4.18}{}$~$\si{\tera\op\per\second}$ on single-device and multi-device,
respectively, demonstrating the highest performance recorded for stencil
programs on FPGAs to date. We then leverage the framework to study a complex
stencil program from a production weather simulation application. Our work
enables productively targeting distributed spatial computing systems with large
stencil programs, and offers insight into architecture characteristics required
for their efficient execution in practice.
\end{abstract}
}

\maketitle
\thispagestyle{plain}
\pagestyle{plain}

\IEEEdisplaynontitleabstractindextext

\section{Introduction}
\label{sec:introduction}
\label{sec:motivation}

Spatial architectures such as FPGAs, Intel's \emph{Configurable Spatial Accelerator}~\cite{csa}, Xilinx' AI Engines~\cite{ai_engine}, and the Cerebras deep neural network accelerator~\cite{cerebras}, are characterized by a large number of small processing units connected by a configurable network.
These systems sacrifice generality and traditional coherence across hierarchical memory subsystems to achieve higher transistor efficiency than load/store architectures (a.k.a. von~Neumann architectures), which is essential to continue scaling in the post-Dennard/Moore era.

Common to spatial architectures is their amenability to be programmed with dataflow abstractions, as this throws away notions of implicit accesses to off-chip resources and communication between parallel processing units, in favor of explicitly programmable off-chip and on-chip data movement.
In this paradigm, computations are laid out spatially on the device, rather than existing as a temporal instruction stream, directly exposing notions of data locality.
The simplest example of this is a pipeline, where each stage synchronously feeds the next.
Systolic arrays add more complexity by extending this to a sequence of synchronous pipelines that communicate in a messaging fashion.
Generally, directed acyclic graphs (DAGs) of pipelines allow arbitrary dataflow, where each node can be attached to multiple producers and consumers.

The temporal locality in iterative or dependent stencil computations is challenging to exploit on load/store architectures, as they require complex tiling schemes~\cite{an5d} and selective fusion of code segments~\cite{modesto, absinthe}.
In contrast, exploiting this reuse via dataflow is intuitive, as consecutive stages can be pipelined and synchronized via their fine-grained dependencies~\cite{sano_multi_fpga}. Implementations of stencils achieving high performance on reconfigurable hardware often assume idealized iterative stencils, as this enables temporal blocking of consecutive timesteps~\cite{zohouri_stencil, soda}, which maps naturally to pipelined architectures.

In this work, we consider the challenging case of arbitrary stencil DAGs, motivated by their existence in numerical climate and weather prediction, where each node is a (potentially complex) stencil operation reading from one or more input memories, and writing its output to one or more consumers.
As a motivating case study, we target an application from the Consortium for Small-scale Modeling (COSMO). The consortium consists of eight national weather services which aim to develop, improve and maintain a non-hydrostatic local area atmospheric model.
The COSMO model is used for both operational \cite{ops1, ops2} and
research \cite{dev1, dev2} applications by the members of the consortium and many universities worldwide.
The stencils used in these simulations are dominated by series of \emph{heterogeneous} stencil computations. Unlike the uniform codes often evaluated in high-performance computing research, these programs run many different stencil operations on many different inputs of varying dimensionality, and exhibit complex dependency patterns between them. 

\begin{figure}[t!]
    \includegraphics[width=\columnwidth]{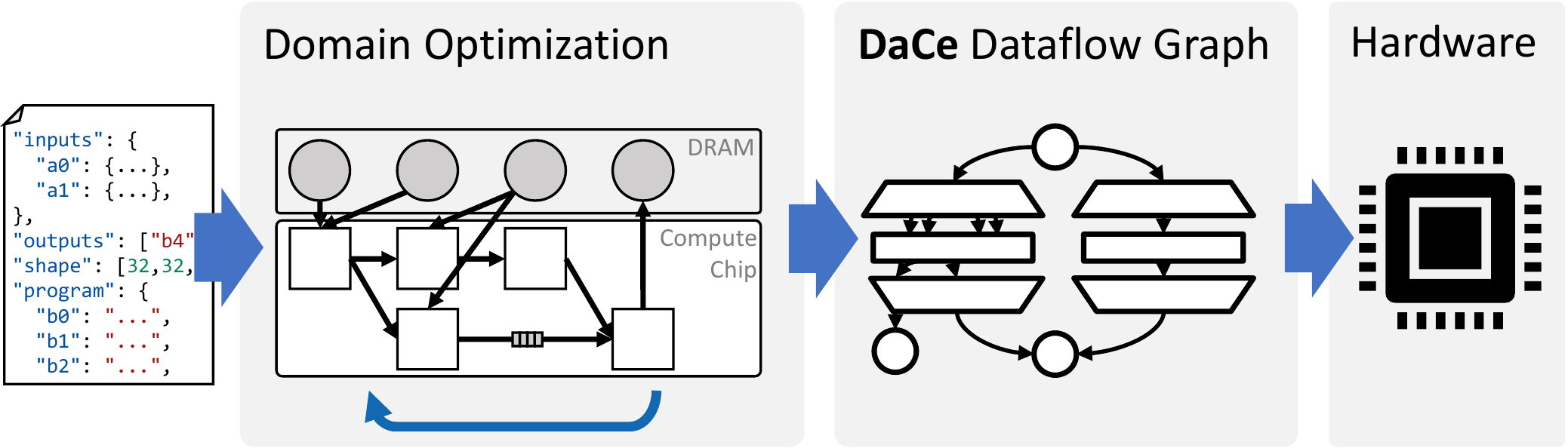}
    \caption{Overview of the StencilFlow end-to-end system.}
    \vspace{-1em}
    \label{fig:posterchild}
\end{figure}

We present a full-stack solution, from a high-level stencil DSL to low-level
spatial program definitions, that are code generated for hardware execution,
summarized in \figureref{posterchild}.
We introduce a method that maps stencil programs to spatial architectures by
using dataflow principles to form compositions that are deadlock free and
maximize the number of active pipelines, based on an analysis of iteration
patterns and the computational source code.
\emph{Fully code-generated} architectures emitted by StencilFlow evaluated on an
FPGA testbed reach $\SI{1.32}{\tera\op\per\second}$ and
$\SI{4.18}{\tera\op\per\second}$ in single-device and multi-device experiments,
respectively, which to the best of our knowledge is the \emph{highest
performance recorded for stencil programs executed on FPGA hardware to date}.
The full source code is available on
GitHub\footnote{\url{https://github.com/spcl/stencilflow}}, exposing productive
high-level Python interfaces, while compiling to highly efficient hardware
through the code-generating backend.

Clear separation of concerns at multiple levels of the stack is a key concept in our approach.
An input program is formulated as a high-level DSL, constraining the program to an analyzable and optimizable form.
Input programs are first optimized on a domain-specific level, where we can perform specialized transformations, such as fusing consecutive stencil nodes.
Then, programs are lowered to a dataflow representation represented in DaCe~\cite{dapp}, where we can control and optimize for data movement.
The dataflow representation is then specialized for the targeted architecture, and finally code-generated to be compiled and synthesized for hardware.
%

\section{Definition of a Stencil Program}
\label{sec:scope}

\begin{figure}[b]
    \centering
    \begin{minipage}{.51\columnwidth}
        \begin{mintedBlock}
            \begin{cpart}{}
{ "inputs": {"a0": {"dtype": "float32",
                    "dims": ["i","j","k"]},
             "a1": {"dtype": "float32",
                    "dims": ["i","j","k"]},
             "a2": {"dtype": "float32",
                    "dims": ["i","k"]} },
  "outputs": ["b4"], "shape": [32, 32, 32],
  "program": {
    "b0": {"code": "a0[i,j,k] + a1[i,j,k]",
           "boundary_condition": {
               "a0": {"type": "constant",
                      "value": 1},
               "a1": {"type": "copy"} } },
    "b1": {"code": "0.5*(b0[i,j,k] + a2[i,k])",
           "boundary_condition": "shrink"},
    "b2": {"code": "0.5*(b0[i,j,k] - a2[i,k])",
           "boundary_condition": "shrink"},
    "b3": {"code": "b1[i-1,j,k] + b1[i+1,j k]",
           "boundary_condition": "shrink"},
    "b4": {"code": "b2[i,j,k] + b3[i,j,k]",
           "boundary_condition": "shrink"} } }
           \end{cpart}
        \end{mintedBlock}
        \vspace{-1em}
        \captionof{listing}{Program description.}
        \label{lst:mapping}
    \end{minipage}\hfill
    \begin{minipage}{.48\columnwidth}
        \vspace{1.55em}
        \begin{minipage}{\columnwidth}
            \includegraphics[width=\columnwidth]{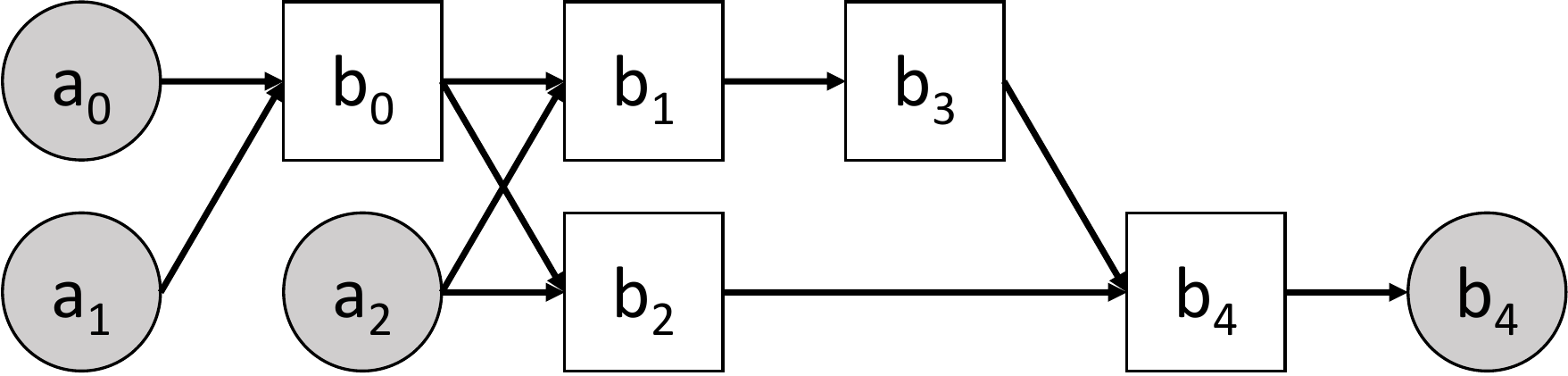}
            \vspace{-1.5em}
            \caption{Corresponding DAG.}
            \label{fig:dag}
        \end{minipage}
        
        \vspace{1.57em}
        
        \begin{minipage}{\columnwidth}
            \includegraphics[width=\columnwidth]{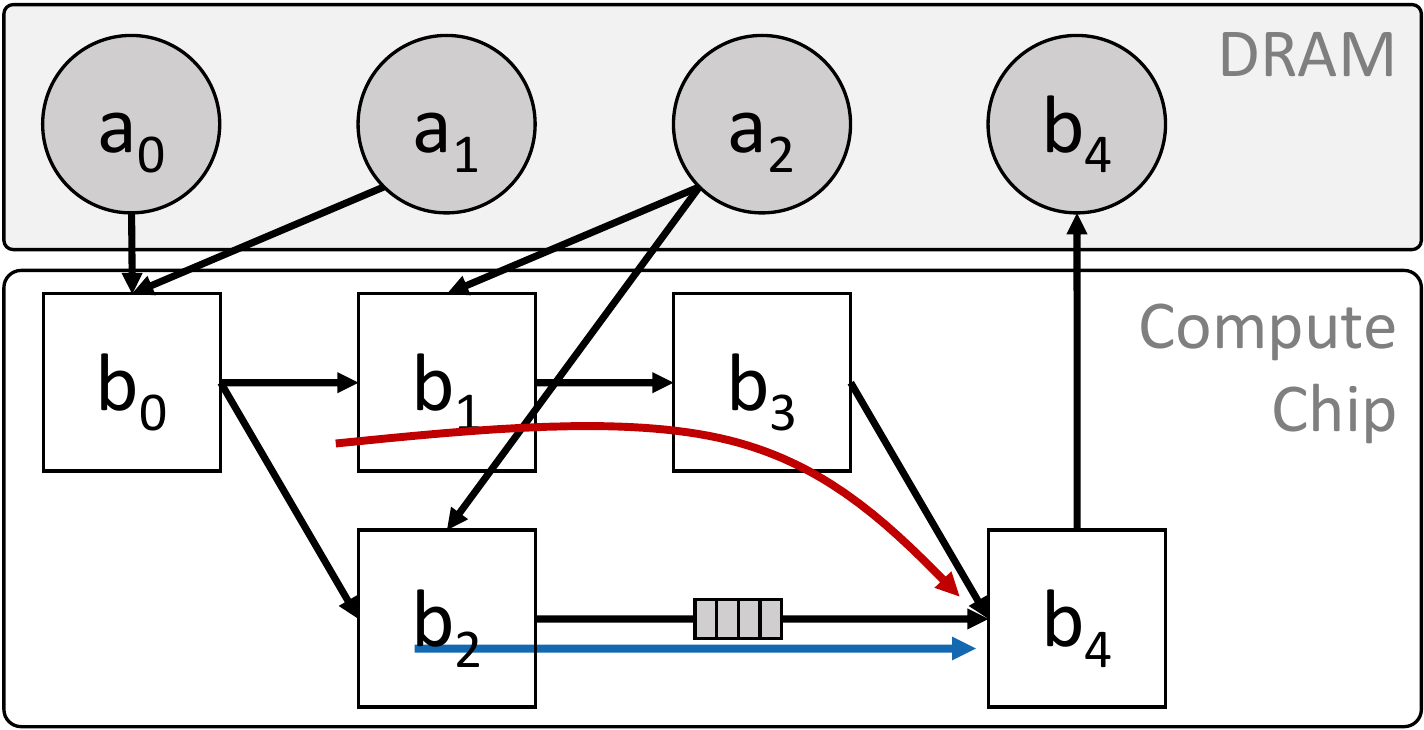}
            \vspace{-1.4em}
            \caption{Hardware mapping.}
            \label{fig:mapping}
        \end{minipage}
    \end{minipage}
    \vspace{-0.2em}
\end{figure}

As the input format of StencilFlow, we define a ``stencil program'' as a
\emph{directed acyclic graph of stencil operations on a structured grid} (an
example is shown in \figureref{dag}), where each node is either a stencil
operation performed on the full output domain or a memory container, and edges
are dependencies between stencils and memories: i.e., outputs produced by one
stencil that are consumed by one or more other stencils, and/or are read
from/written to memory. Each stencil takes one or more inputs, that are sourced
either from off-chip memory, or fed by a previous stencil evaluation, and
produces exactly one output. To support a broader class of computations present
in weather models considered, we furthermore allow stencils to read from
lower-dimensional inputs: e.g., a 3D stencil can read from a 2D, 1D, or even
``0D'' (scalar) arrays using subsets of its indices.
A stencil node is defined by:
\begin{itemize}[itemsep=0pt, leftmargin=*]
    \item A definition of each logical input that is read, which we refer to as ``fields'', with a corresponding data type, and a sequence of offsets relative to the center (``field accesses'').
    \item A code segment describing the computation at each point in the
    iteration space, where only the specified input accesses (including 0D
    constants) can be used in computations. Since it is important to know the latency of computations, the code is restricted to be \emph{analyzable} (i.e., no external data structures or external functions, with the exception of standard math functions). However, ternary functions/conditionals are allowed, \textbf{including data-dependent branches}.
    \item A series of boundary conditions, defining how out-of-bounds accesses should be handled.
\end{itemize}
Currently supported boundary conditions include: \textbf{constant}, where out of bounds accesses are replaced with a given constant value; \textbf{copy}, where out of bounds accesses are placed by the value at offset 0 in all dimensions (the ``center'' value); and \textbf{shrink}, where all computed values that read out of bounds values are simply ignored in the output. The former two are specified per input, whereas shrink is specified on the output.

To facilitate productive definition of stencil programs, we define a simple JSON-based input format, which only requires the minimum amount of information necessary to instantiate the stencil DAG to be specified explicitly. An example is shown in \listingref{mapping}. In practice, the definition must additionally provide data sources for each input field. Stencil programs can have 1, 2, or 3 dimensions, but assume all stencils iterate over the same iteration space (although they can have variable constant offsets into the output field).

\section{Mapping to Distributed Hardware}
\label{sec:mapping_to_distributed_hardware}

There is a substantial body of previous work on mapping single stencil operations to reconfigurable hardware~\cite{zohouri_stencil, sano_multi_fpga, fu_clapp, soda}, where high performance is achieved by chaining many consecutive timesteps together as a rich source of temporal locality. Some of this methodology carries over to the more general scenario we consider here, but we must additionally consider forks and joins in the stencil program, inputs and outputs shared by multiple producers and consumers, heterogeneity and complexity in stencil computations, and mapping the graph to multiple devices.

\subsection{Mapping to Hardware}
\label{sec:mapping_to_hardware}

For our hardware mapping, \emph{we work from the base assumption that every stencil operation in the dependency graph is mapped to simultaneous dedicated logic} (stencil \emph{units}/operators), even if this requires the design to span multiple devices.
All stencil operations are scheduled simultaneously, operating in a \textbf{fully pipeline parallel manner}. In this scenario, production and consumption rates are identical across the dataflow graph, allowing the runtime to be modeled as a single, deep pipeline (described in \secref{performance_model}).

Each stencil unit executes a pipeline, which processing a number of cells equal to the product of the input dimensions, where logic required to handle out-of-bound accesses is predicated into the pipeline. The next cell is evaluated as soon as all inputs required \emph{for that cell} are ready. This way, all dependencies between stencils become fine-grained on a per-cell level. This spatial computing view is distinct from the load/store view, as we default to \emph{perfect data reuse} (i.e., we exploit all available temporal locality). In contrast, the efficiency of computations on load/store architectures relies on maintaining a task granularity suitable for the architecture (large number of identical threads on GPU, small number of large tasks on CPU). Kernel fusion is thus a critical optimization to achieve the right task granularity~\cite{absinthe} for performance through spatial locality, whereas \emph{StencilFlow programs are executed in a fully ``fused'' schedule}, but are instead concerned with satisfying the off-chip and on-chip memory constraints (fusing stencil operators takes a different meaning, described in \secref{stencil_fusion}).

Inputs are provided to each stencil unit through on-chip channels with a
\emph{compile-time fixed size}, where the producer can either be another stencil
unit (i.e., a dependency), or a memory unit reading directly from off-chip
memory. If one or more inputs are not ready, the pipeline must stall while
waiting for the remaining inputs to arrive. For any DAG that is not a
multi-tree, this can result in deadlocks if channel capacities are insufficient
to buffer inputs ready early until inputs ready later arrive, due
to the circular dependency implied by each data exchange requiring the
receiver and sender to not be \emph{full} and not be \emph{empty}, respectively.
We must thus take all paths through the DAG into account when deciding the size
of buffers between dependencies.

\begin{figure}[hbt]
  \centering
  \includegraphics[width=.9\columnwidth]{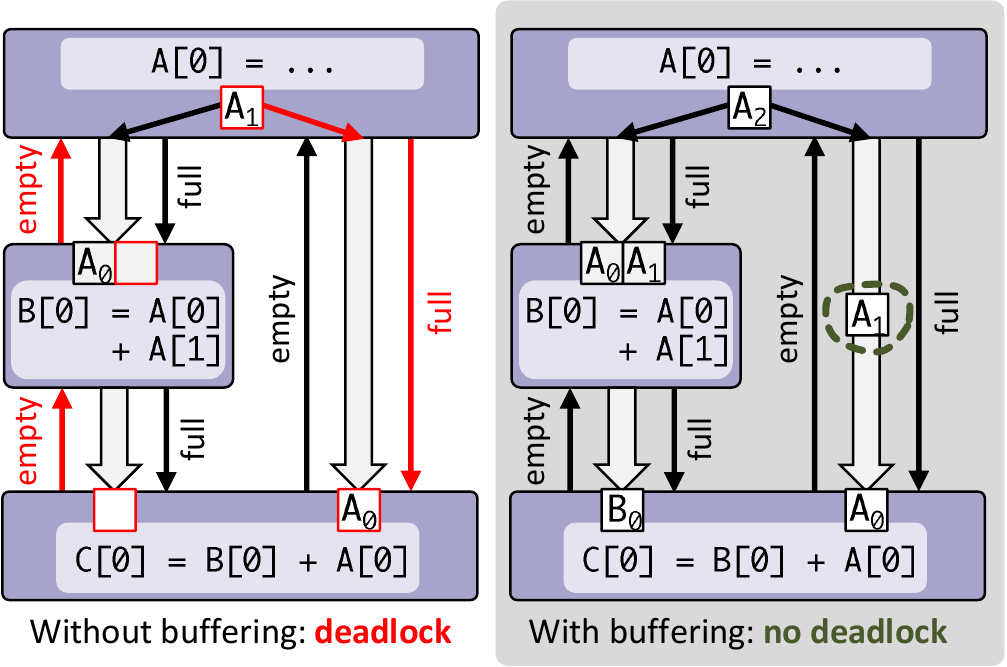}
  \caption{Preventing deadlocks by injecting buffers.}
  \label{fig:deadlock}
\end{figure}

In the example shown in \figureref{deadlock}, the stencil unit computing
\texttt{C} requires data from both stencil units \texttt{A} and \texttt{B} to
begin streaming. The results streamed out of \texttt{A} are also required by
\texttt{B}. On the left hand side, \texttt{C} is waiting for data from
\texttt{B} (i.e., for the data stream to not be empty), \texttt{B} is waiting
for additional data from \texttt{A}, and \texttt{A} is waiting for \texttt{C} to
accept the data (i.e., for the data stream to not be full), thus forming a
circular dependency. Without additional buffering, this results in a deadlock.
By adding an appropriate buffer between \texttt{A} and \texttt{C} (right hand
side), we can inject sufficient credits to tolerate the delay induced by the
path through \texttt{B}. We describe how StencilFlow computes the buffer depths
required to prevent deadlocks and ensure continuous streaming operation in
\secref{delay_buffering}.

\subsection{Mapping to the Distributed Setting}
\label{sec:mapping_to_distributed_setting}

\begin{figure}[tb]
    \centering
    \includegraphics[width=\columnwidth]{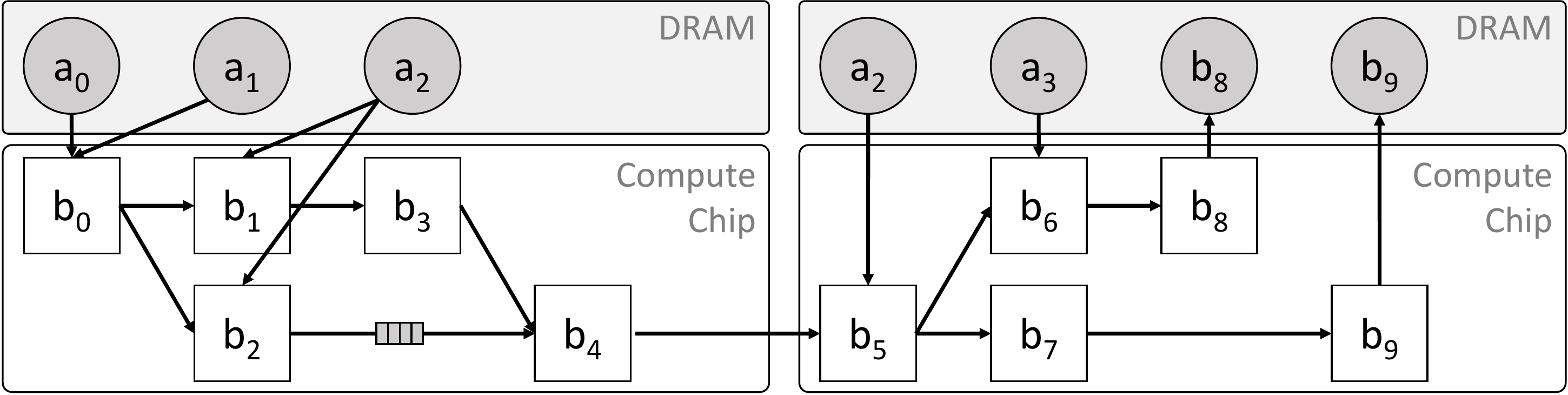}
    \caption{Stencil program spanning two devices.}
    \label{fig:multi_device}
\end{figure}

To scale beyond the off-chip memory bandwidth, on-chip memory capacity, and logic resources available on a single chip, we let designs scale to multiple devices. For modeling and code generation, this means that certain inter-stencil connections will cross devices, and thus imply communication across the network. Furthermore, data located in off-chip memory must be present on any device that accesses it, implying potential replication to multiple devices that require it. In the example shown in \figureref{multi_device}, $a_2$ is accessed by stencils on either device, requiring it to exist in both DRAM memories.

To implement inter-node communication in practice, we leverage the Streaming Message Interface~\cite{smi} (SMI), which exposes communication as channels with FIFO semantics, resulting in inter-node communication being nearly identical to intra-device communication between stencils in the generated code. With the target in mind, the following will describe the program analysis, and the central components of the StencilFlow stack, required to build these spatial architectures.

\section{From DAG to Dataflow}
\label{sec:dag_to_dataflow}

The StencilFlow framework analyses the stencil DAG, and uses this to construct a dataflow graph that maps to efficient hardware. Data reuse happens both internally in each stencil, facilitated by ``internal buffers'', and on the edges between stencil nodes, referred to as ``delay buffers''.

\subsection{Internal Buffers for Intra-Stencil Reuse}
\label{sec:internal_buffers}

\begin{figure}[b]
    \centering
    \begin{minipage}[b]{.49\columnwidth}
        \centering
        \includegraphics[page=1,width=.82\columnwidth]{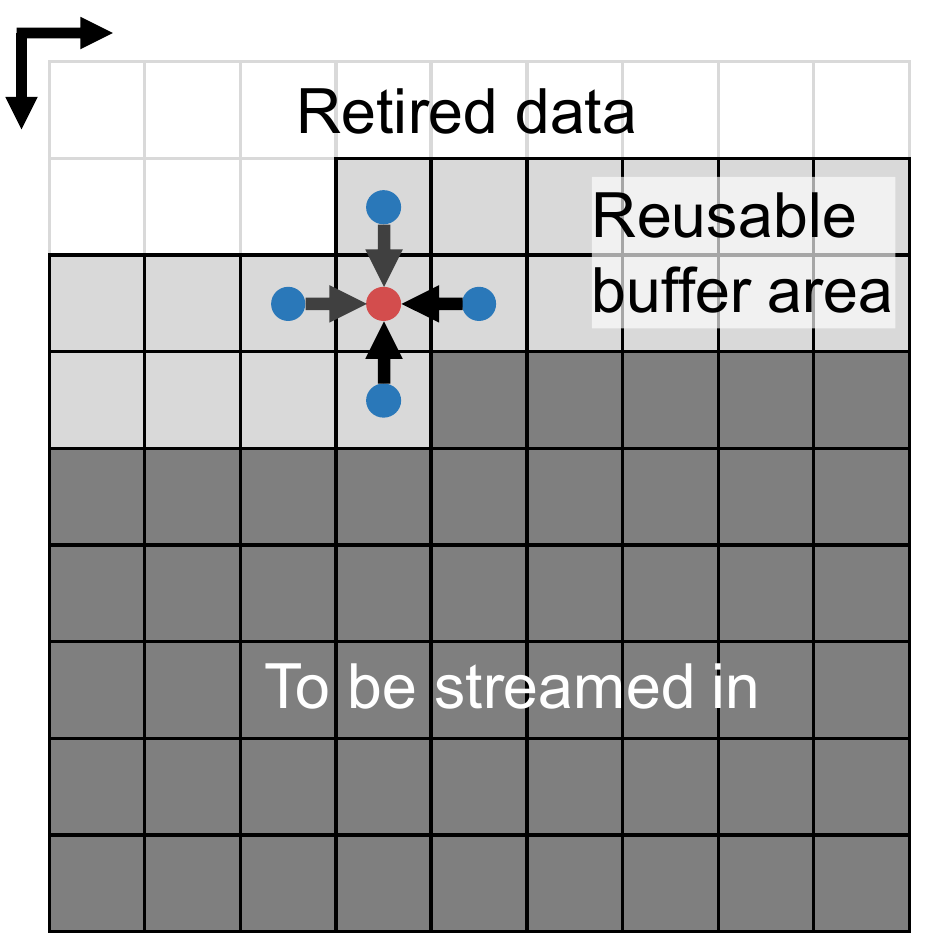}
        \caption{Internal data reuse buffer.}
        \label{fig:internal_buffer}
    \end{minipage}\hfill%
    \begin{minipage}[b]{.49\columnwidth}
        \includegraphics[width=\columnwidth]{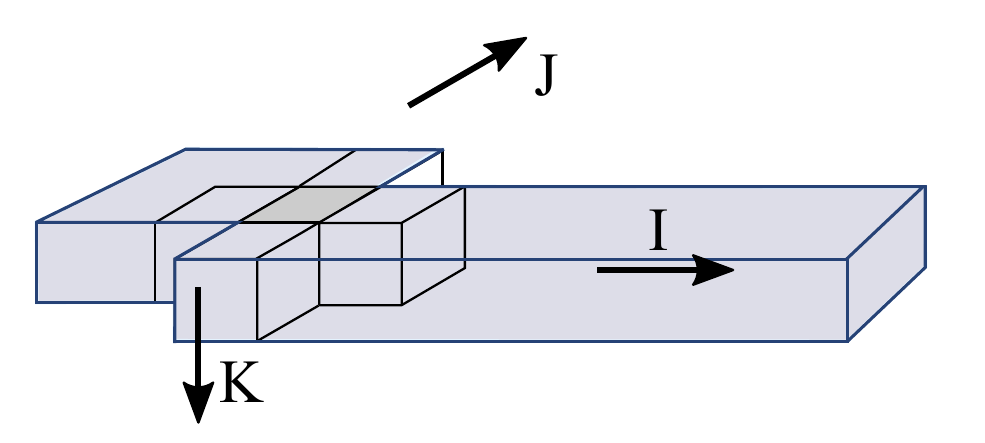}\\
        \includegraphics[width=\columnwidth]{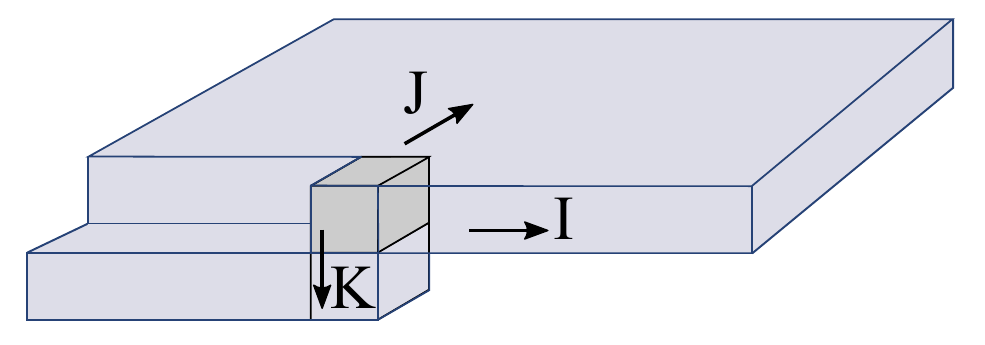}
        \caption{Varying buffer shapes.}
        \label{fig:buffer_shapes}
    \end{minipage}
\end{figure}

The most straightforward source of temporal locality comes from within each
stencil operation, where the same input field is often accessed at multiple
offsets relative to the center, illustrated in \figureref{internal_buffer} for
accesses $\{[-1, 0],[0, -1], [0, 1], [1, 0]\}$ in a 2D iteration space.
Furthermore, in the global dataflow setting, the core assumption of StencilFlow
is that data should only be loaded once, streaming directly between kernels
without going through off-chip memory.

A stencil node has 0 or 1 internal buffers per field accessed, depending on whether there are multiple accesses to the given field within the stencil. The size of each buffer is determined by the \emph{largest distance between any two offsets in memory order, plus one} (or plus the vector width, in the case of vectorized kernels) in the stencil iteration space: e.g., in a 3D iteration space of shape $\{K, J, I\}$, two accesses \texttt{a[0, 1, 0]} and \texttt{a[0, -1, 0]} require buffering two 1D rows ($2 I + W$ elements, where $W$ is the vector width), while two accesses \texttt{b[0, 0, 0]} and \texttt{b[1, 0, 0]} require buffering a 2D slice ($2 I J + W$), shown in \figureref{buffer_shapes} top and bottom, respectively. In general, buffers sizes can be up to a constant number of $(D-1)$-dimensional slices for a $D$-dimensional stencil.

StencilFlow computes the internal buffer size for each field, for each stencil, independently. However, the schedule for when the pipeline starts writing each buffer is dependent on the other fields accessed. For example, if a stencil reads multiple fields with internal buffer sizes $\{B_1,\dotsc,B_F\}$, each internal buffer can be only start to be filled after the first $B_i - \max\{B_1,\dotsc,B_F\}$ iterations (the largest buffer(s) will always start reading immediately), so it is synchronized with the other fields. Additional accesses \emph{in between} the ``highest'' and ``lowest'' offset in memory order do not affect the total buffer size, although they can affect the buffer implementation in practice by adding more parallel accesses into the buffer.

Filling the internal buffers also affects the latency, and transitively the runtime, of the stencil program. A stencil node cannot begin computations before all operands are available, which only happens once all internal buffers have been filled. As the size of buffers is exactly the distance between the lowest and highest accessed index in order of the stencil iteration space, the \emph{initialization phase} of a stencil is given by $\max\{B_1,\dotsc,B_F\}$, which is crucial to the delay buffer calculation described in the following.

\subsection{Delay Buffers for Inter-Stencil Reuse}
\label{sec:delay_buffers}
\label{sec:delay_buffering}

Edges between stencils in the DAG enable data reuse by replacing expensive
round-trips to off-chip memory with direct dataflow. Furthermore, if multiple
stencils require data from the same input field, it is sufficient to read it
from memory once, and stream the data to all stencils requiring it. StencilFlow
exploits all such opportunities, while preventing the deadlock scenario
illustrated in \figureref{deadlock}. This requires synchronizing inputs to
consumers by adding buffers that delay the data (i.e., inject sufficient
credits) until all inputs are ready without blocking the producer(s). We
annotate these delay buffers on edges in the dataflow graph, corresponding to
FIFO channel depths.

\begin{figure}[b]
    \centering
    \includegraphics[page=2,width=.6\columnwidth]{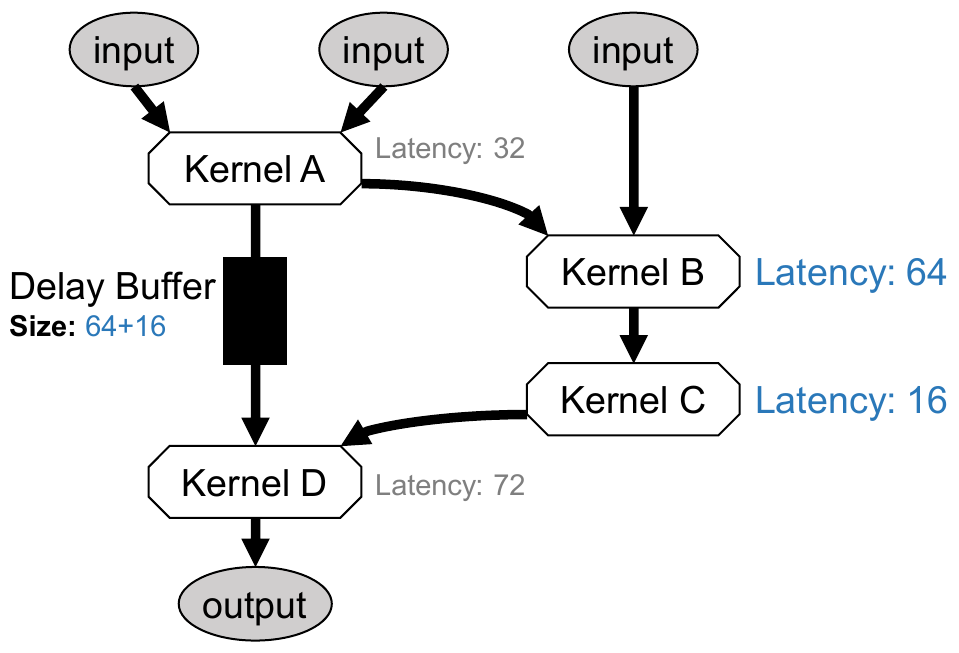}
    \caption{Delay buffers on edges enable reuse and deadlock freedom.}
    \label{fig:delay_buffers}
\end{figure}

There are two factors that determine delays in the DAG. First, the AST formed by
computation of a stencil operation forms another DAG, whose critical path adds a
delay between a sequence of inputs entering and exiting the pipeline. Computing
the critical path requires latency information for each operation performed,
which is both type and architecture dependent. As a result, these latencies can
be provided as configuration to the framework, and default to conservative
values to account for the worst case scenario. We note that these delays are
typically small (${<}100$ cycles), and do not contribute significantly to the
overall fast memory usage, even when conservatively overestimated. More
importantly, delays occur in the initialization phases within each stencil,
where internal buffers are being filled before enough data is available to start
computations. Each stencil node in the stencil program will contribute
$\max\{B_1,\dotsc,B_F\}$ elements to this delay, where $\{B_1,\dotsc,B_F\}$ is
the set of $F$ internal buffer sizes for the given stencil.

To determine the size of delay buffers on the edges arriving at a given node, we
traverse the DAG backwards from the node, computing the latency contributions
along all possible paths, from all possible source nodes, and for each edge,
\emph{including the contribution of the initialization phase of the node
itself}, recording the highest delay encountered per edge. The buffer size on
each edge is then the highest delay found for that edge, subtracted from the
highest delay found across \emph{all} edges (it follows that each node will have
at least one incoming edge with delay size zero). Similar to internal buffers,
the maximum size of delay buffers is proportional to the size of a
$(D-1)$-dimensional slice of the iteration space. An example of annotated delay
buffers in shown in \figureref{delay_buffers}.

\subsection{Vectorization}
\label{sec:vectorization}

When insufficient reuse is present in a target program, we can employ vectorization to increase parallelism and memory bandwidth utilization, in order to approach a compute logic or memory bandwidth bound. To this end, we allow StencilFlow input programs to specify a vectorization factor, which will not only affect the generated hardware, but also the dataflow analysis. Vectorizing by a factor of $W$ reduces the number of iterations in the inner loop of all stencils in the program by a factor of $W$, which affects the size of initialization phases, and transitively the delay buffers in the system. In addition to directly increasing the bandwidth requirement and parallelism in the program, vectorization can also have the subtler effect of coarsening stencil nodes, increasing the ratio of ``useful'' compute logic to overhead logic. We can thus also use vectorization in time tiling-like scenarios to coarsen simple stencils and increase the achievable performance.\\[0.5em]
Once the stencil program has been enriched with the appropriate internal buffer and delay buffer sizes, the resulting graph is emitted to the data-centric backend for domain-specific and low-level optimizations.

\section{Data-Centric Abstract Representation}
\label{sec:dace}

We use the Data-Centric (DaCe)~\cite{dapp} framework as a dataflow representation and backend for the hardware mapping. 
DaCe defines a graph-based development workflow that maintains a separation of concerns between domain scientists and performance engineers, based on the observation that the vast majority of hardware optimizations are centered around data movement reduction. 
DaCe generates high-performance code for both load/store architectures and reconfigurable hardware, and supports high-level synthesis (HLS) backends for both Xilinx and Intel FPGA architectures.

\begin{figure}[t]
    \centering
    \includegraphics[width=.9\columnwidth]{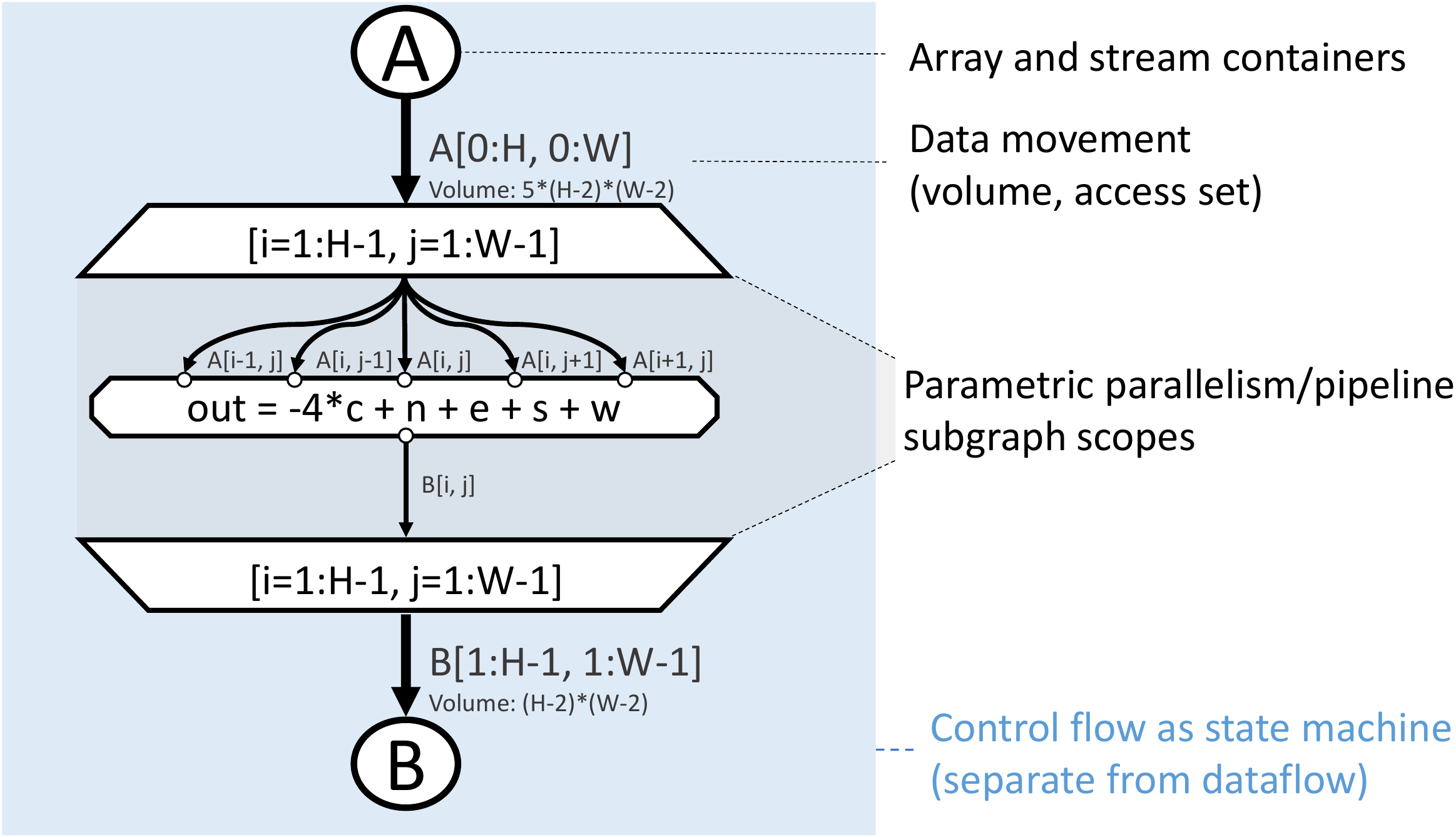}
    \caption{Data-centric representation (SDFG) of a Laplace stencil.}
    \label{fig:sdfg}
\end{figure}

DaCe separates program definition from its optimization by using the the
Stateful DataFlow multiGraph (SDFG) representation. In each multigraph, data
movement (edges) is explicitly separated from data/FIFO containers and
computations (nodes). These acyclic multigraphs are, in turn, nested within
state machines (directed graphs) that represent the control flow of the
application.  An example of a two-dimensional Laplace operator is shown in
Fig.~\ref{fig:sdfg}.

In the DaCe workflow, a program is developed in a frontend language (e.g., Python, StencilFlow) by a domain scientist. All subsequent hardware mapping and optimizations are performed on the SDFG separately, by a performance engineer. Optimization and hardware mapping is achieved via graph rewriting rules, called \textit{transformations}, for data movement reshaping, scheduling parallel subgraph scopes to processors, and modifying data storage/layout. Transformations are user-extensible and written in Python interfaces, allowing both domain-specific and general purpose optimizations, and enabling knowledge transfer between applications. Adaptations to the graph are saved separately from the source code, allowing the original source code to be modified without changing the optimization scheme.

\subsection{Extensions to DaCe}
\label{sec:dace_extensions}

To support this work, we extend the DaCe framework, introducing a new type of dataflow node, pipelined scopes, and three new transformations.
In particular, we extend the SDFG with the concept of \textit{library nodes}.
Library nodes function similarly to computational nodes, but encode domain-specific information and contain multiple implementation targets, which translate into different subgraphs upon expansion.
The StencilFlow-specific library node \texttt{Stencil} was developed for this
work, and will be used extensively throughout the following.
Since the high-level semantics of library node types are known, they allow performance engineers to develop domain-specific transformations, such as algebraic contractions (e.g., double transposition) and others. With library node expansions potentially containing other library nodes, multi-level coarsening and transformations are thus enabled in SDFGs, inspired by the MLIR~\cite{mlir} stack.

As a useful shorthand for pipelined iteration spaces, we introduce the \emph{pipeline scope}, augmented with information on initialization and draining phases, to easily allow the programmer to inject specialized behavior during initialization, streaming, and draining phases. For StencilFlow, this allows encoding the internal buffer initialization phase, and draining phases where results are still being computed only using data present in local buffers, thus omitting reads from inputs.

\begin{figure}[b]
    \centering
    \includegraphics[width=.9\columnwidth]{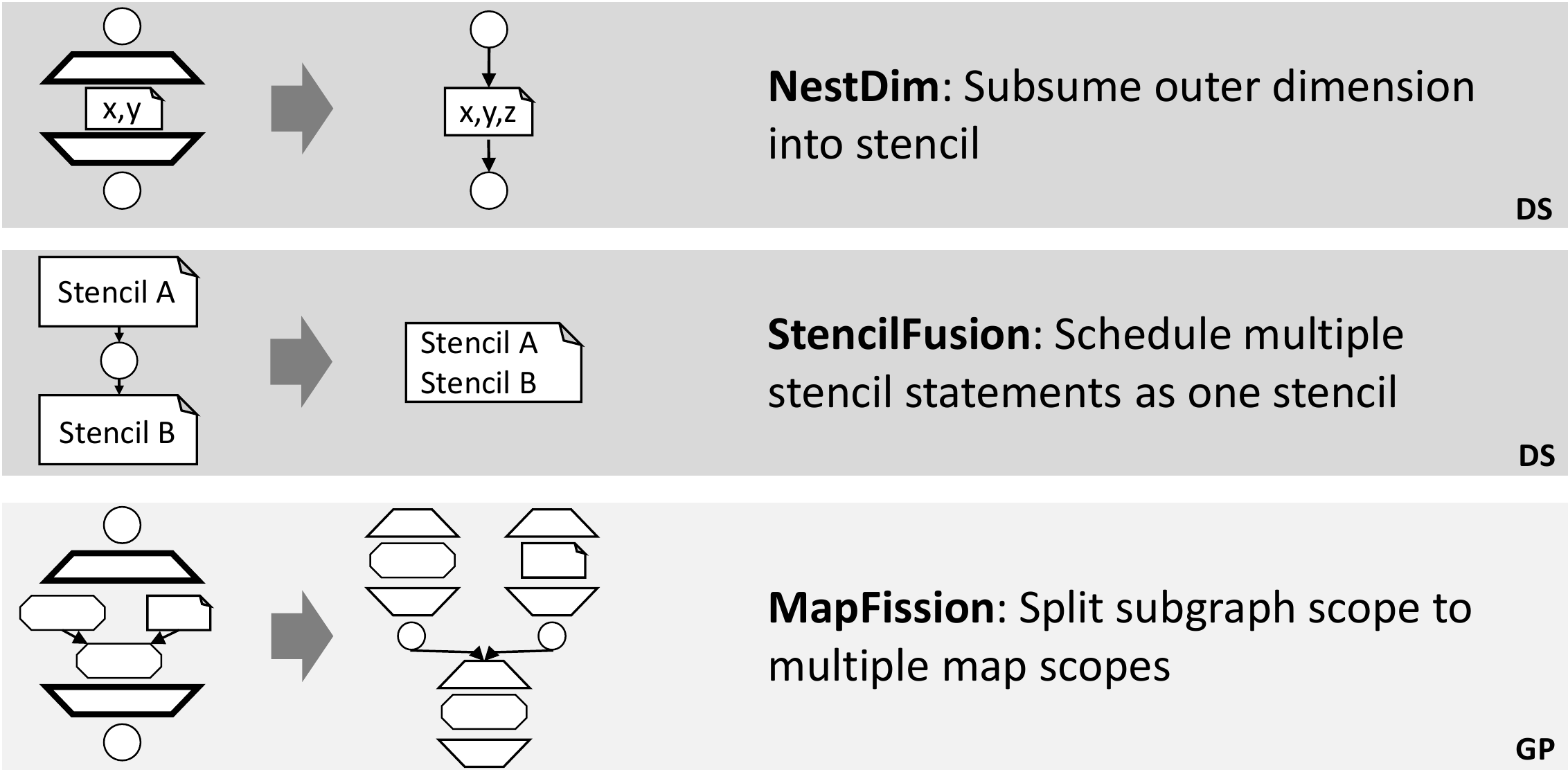}
    \caption{Transformations used.\\(DS: Domain-Specific, GP: General-Purpose).}
    \label{fig:xforms}
\end{figure}

With the domain-specific concepts enabled by library nodes, we are now able to develop transformations for stencil programs on reconfigurable hardware. We develop both domain-specific and a general-purpose transformation, summarized in Fig.~\ref{fig:xforms}. 
\texttt{NestDim} reschedules stencil computations by taking multiple, parametrically-parallel stencils and creating one stencil, which can be mapped into different schedules on hardware. \texttt{StencilFusion} schedules multiple dependent stencils as one stencil with multiple statements, differing from standard map fusion by taking boundary conditions and redundancy into account. For general-purpose transformations, we add the \texttt{MapFission} transformation, which splits a parallel subgraph into multiple parallel subgraphs (which can in turn be rescheduled), introducing temporary storage between the subgraph components. The \texttt{NestDim} and \texttt{MapFission} transformations are used as a tool to extract stencil programs from existing SDFGs to analyze them in StencilFlow, while \texttt{StencilFusion} is an optimization for both load/store and spatial architectures, described in the context of StencilFlow below.

\subsection{Spatial Stencil Fusion}
\label{sec:stencil_fusion}

\begin{figure}
    \begin{subfigure}{.48\columnwidth}
        \centering
        \includegraphics[height=8em]{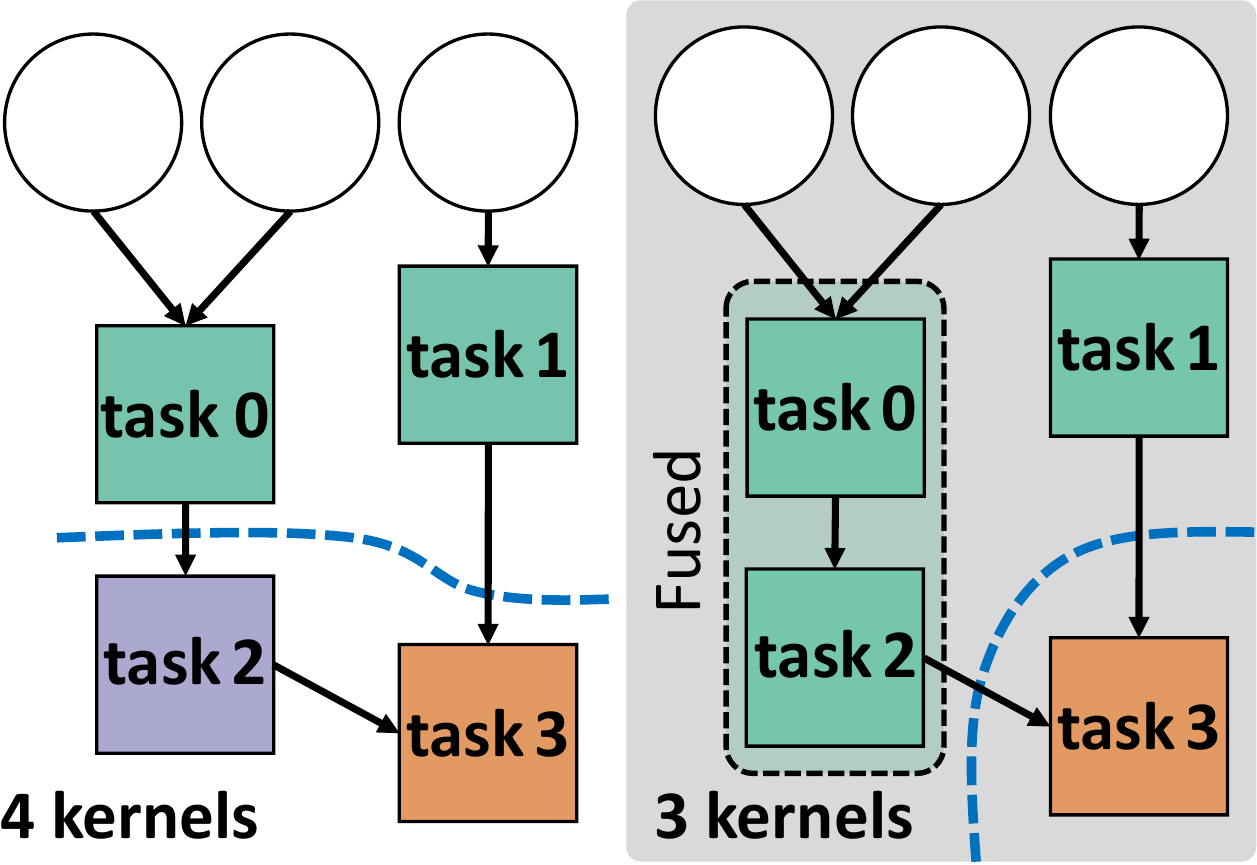}
        \caption{\small Load/store stencil fusion.}
        \label{fig:fusion_cpu}
    \end{subfigure}\hfill%
    \begin{subfigure}{.48\columnwidth}
        \centering
        \includegraphics[height=8em]{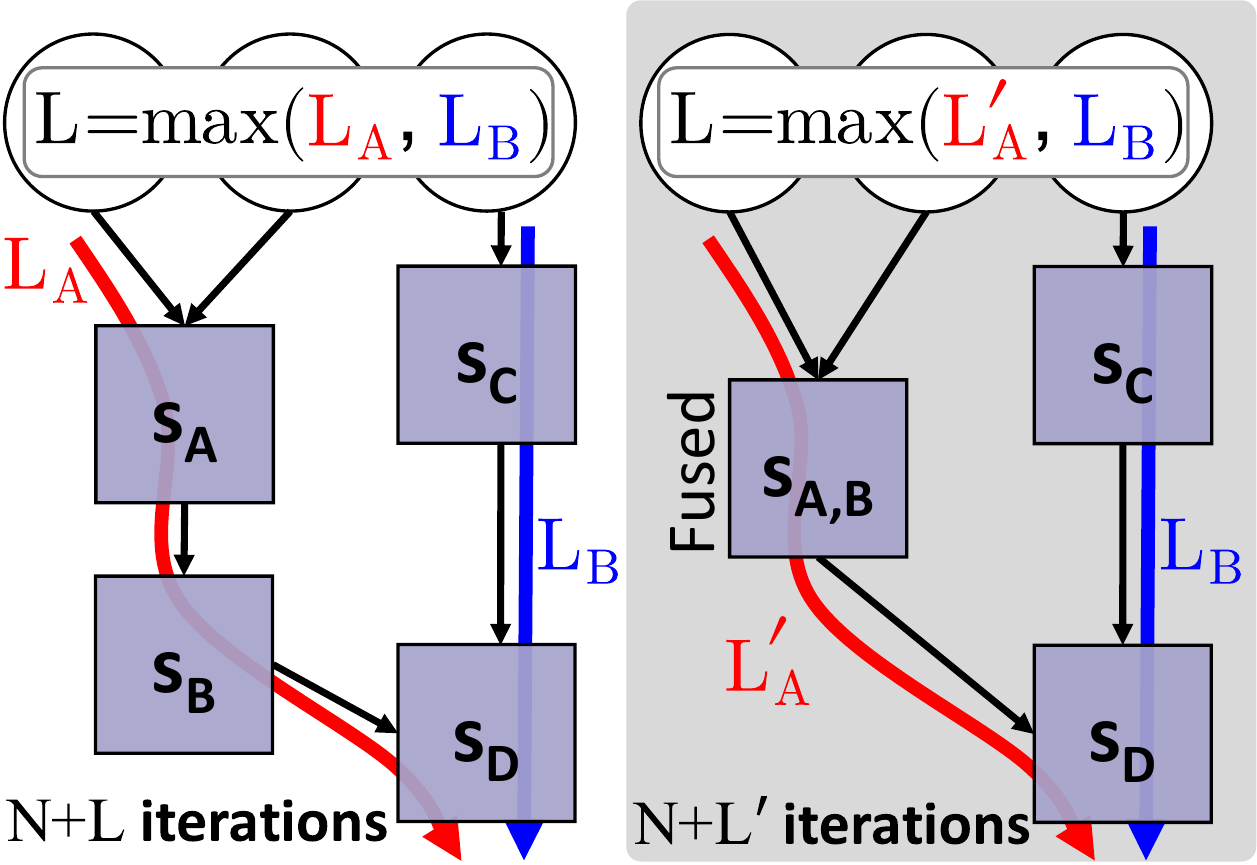}
        \caption{\small Stencil fusion in StencilFlow.}
        \label{fig:fusion_spatial}
    \end{subfigure}
    \caption{Unlike load/store fusion, spatial fusion only reduces latency.}
    \label{fig:fusion}
\end{figure}

On load/store architectures, fusing consecutive stencils is used to increase performance by improving data locality, reducing write/read roundtrips from off-chip memory, and reducing context scheduling overhead~\cite{serial_combinators}. When applying the transformation on StencilFlow dataflow graphs, the effect is somewhat different, as the schedule of the spatial architecture is already fully ``fused'' into a global pipeline. Instead, fusing stencils has the following effects:
\begin{itemize}[itemsep=0pt, leftmargin=*]
    \item The critical path through the program can be reduced by combining the initialization phases (see \secref{internal_buffers}) of two consecutive stencils.
    \item Internal buffers for the same input field are combined into a single internal buffer.
    \item Multiple smaller delay buffers can be combined into fewer, larger buffers, which affects hardware utilization, depending on the granularity of on-chip memory on the target platform.
    \item Combined code sections increase the opportunity for common subexpression elimination by the optimizing compiler.
    \item Coarser stencil nodes increase the ratio of ``useful'' logic to the number of pipelines instantiated, which can affect spatial resource overhead.
\end{itemize}
The difference between fusing tasks on load/store architectures and the spatial fusion performed here is illustrated in \figureref{fusion}. On load/store architectures (\figureref{fusion_cpu}), the total number of scheduled kernels is reduced when fusing task~0 and task~2 into a single kernel. In \figureref{fusion_spatial}, all operators are already scheduled in parallel, but the initialization latency can be reduced \emph{if the fused nodes $s_A$ and $s_B$ are on the critical path}.

In our dataflow canonicalization pass, we define a collection of heuristics for
fusing two stencils so that these effects are observed. Firstly, the necessary
conditions for fusion are checked, namely that the two stencils operate on the
same data shape (correlating to iteration space) and that they have the same
StencilFlow boundary condition definitions. Then, we only consider stencils that
are connected by one data container node $u$ with $\deg (u){=}2$, in order to
ensure that all stencils (fused or otherwise) have a single output. Finally, we
ensure no other instances of $u$ exist in other states, so that it can be
completely removed from the graph without adding an extra write to off-chip
memory.

For the experiments in this work, we perform aggressive stencil fusion of input programs, as this is observed to reduce overall logic through the coarsening of stencil nodes, and slightly reduces runtime by pruning initialization latencies.

\section{Code Generation}
\label{sec:code_generation}

StencilFlow relies on DaCe backends to generate the final kernel code, which is
passed to an optimizing compiler. For the experiments performed in this work, we
target the Intel FPGA SDK for OpenCL backend~\cite{altera_opencl}, which is a
high-level synthesis (HLS) compiler, emitting RTL code from annotated OpenCL.
Being an HPC-oriented framework, DaCe automatically performs necessary
annotations for pipelining, unrolling, and coalescing loops emitted from
parametric maps in the dataflow graph, splits parallel sections into processing
elements (i.e., OpenCL kernels), annotates buffer depth properties for channels,
declares kernels as \texttt{autorun} when possible, inlines constants, and
performs conversions between vectorized and non-vectorized data types. Host code
necessary to interface with the kernel and the necessary memory copies are
generated, and the final program can be called by using the high-level Python
interface. By using DaCe for code generation and by using the library node
abstraction for stencil computations, supporting Xilinx FPGAs, emitting RTL code
directly, or targeting other spatial systems entirely will only require
adapting the stencil library node expansion, provided that support for the
desired architecture is present in/added to the DaCe framework. 

\subsection{Intel FPGA Optimizations in DaCe}
\label{sec:intel_fpga}

When targeting the Intel HLS compiler, delay buffers are represented as DaCe
streams with a given buffer size, which are mapped to the Intel OpenCL
\texttt{channel} abstraction, that in turn are mapped to FIFOs in hardware. We
target the shift register pattern in Intel's OpenCL compiler to efficiently
implement internal buffers within each stencil node. To achieve this in DaCe, a
data container spanning the full width of each internal buffer is created,
injecting and shifting elements every cycle. ``Tap'' points (constant offset
accesses) into the array are then connected to the stencil, where offsets are
generated from the distance between accesses flattened into a 1D iteration
space.

\begin{figure}[b]
    \centering
    \includegraphics[width=\linewidth]{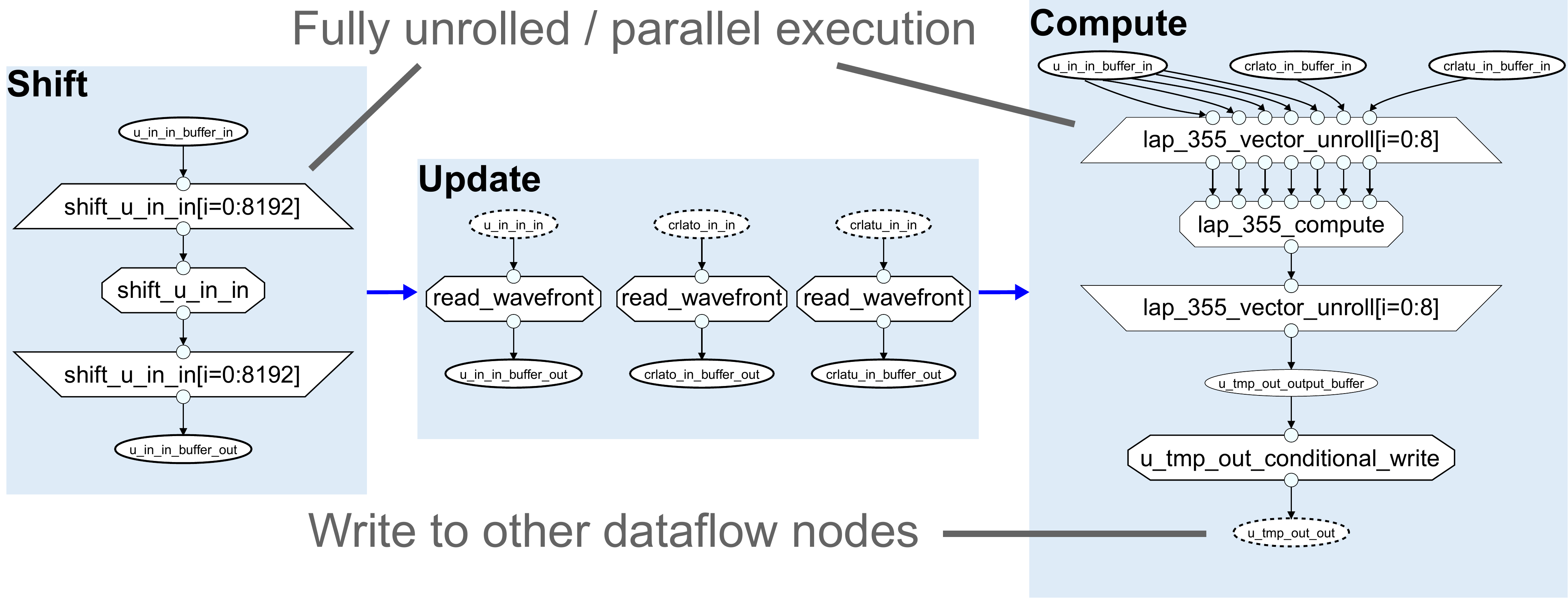}
    \caption{Data movement occuring per iteration of a stencil.}
    \label{fig:nested_sdfg}
\end{figure}

The processing done per cell in an expanded stencil library node is shown in \figureref{nested_sdfg}. The graph contains three consecutive components: a \textbf{shift} phase, containing a fully unrolled map scope (trapezoids) where a ``tasklet'' (octagon) shifts each entry of the shift register memory (ovals) by the vectorization width to $i{+}W$; an \textbf{update} phase, where new values are read from the input channel (dashed border) into the front of each shift register by a tasklet; and a \textbf{compute} phase, where the buffers are accessed at all tap points and fed to the main computation tasklet, which is parametrically unrolled to treat each element in the vector with potentially different boundary conditions, and passes through another tasklet that conditionally writes the output stream if the stencil is not in the initialization phase. This full graph will be wrapped in a parametric scope that defines the iteration space of the stencil program, which is fully pipelined, such that all three phases are executed in a pipeline parallel manner. The input and output streams (dashed borders) are connected to the appropriate producers and consumers in the global dataflow graph.
Source nodes are instantiated as dedicated prefetchers that can read ahead of computations, and dedicated writers are instantiated at sink nodes that can buffer data while waiting for DRAM writes.

\subsection{Generating Distributed Programs}
\label{sec:SMI}
\label{sec:smi}

To code-generate distributed implementations, we integrate an OpenCL implementation of the \emph{Streaming Message Interface}~\cite{smi} into the DaCe backend. SMI extends HLS with a distributed memory programming model for reconfigurable hardware that unifies message passing with pipelined stream-based communication of data, such that cross-chip communication is expressed the same way as on-chip communication.

When a stencil program spans multiple devices, the computation running on each device is represented by a separate DaCe program, as it will compile to separate bitstreams that must be configured to each device in the sequence.
Devices communicate via \emph{remote streams}, which are DaCe streams annotated with having a source/destination located on a different device, which will trigger the SMI backend to code generate the relevant networking code and emit streaming message communication.

If multiple network connections are present between two endpoints, SMI can split a communication stream into two or more substreams following different channels across the network, and recombine them at the other end, allowing for a multiplicative increase in achievable bandwidth. In StencilFlow, we exploit this to increase the vectorization width and number of channels spanning across devices.

\subsection{Reference Code}
\label{sec:reference_code}

Using domain-specific stencil nodes within the DaCe framework, we are able to
maintain a high-level view of the program, which enables optimization for
different architectures. While exploring CPU and GPU performance is out of the
scope of this work, we exploit this capability to generate reference
CPU-executed graphs where stencil evaluations are executed sequentially in
topological order (i.e, no fusion or parallelism between stencil evaluations),
which we can verify against the generated hardware kernels.

\section{Workflow and Artifacts}
\label{sec:workflow}

\begin{figure}
    \centering
    \includegraphics[width=.9\columnwidth]{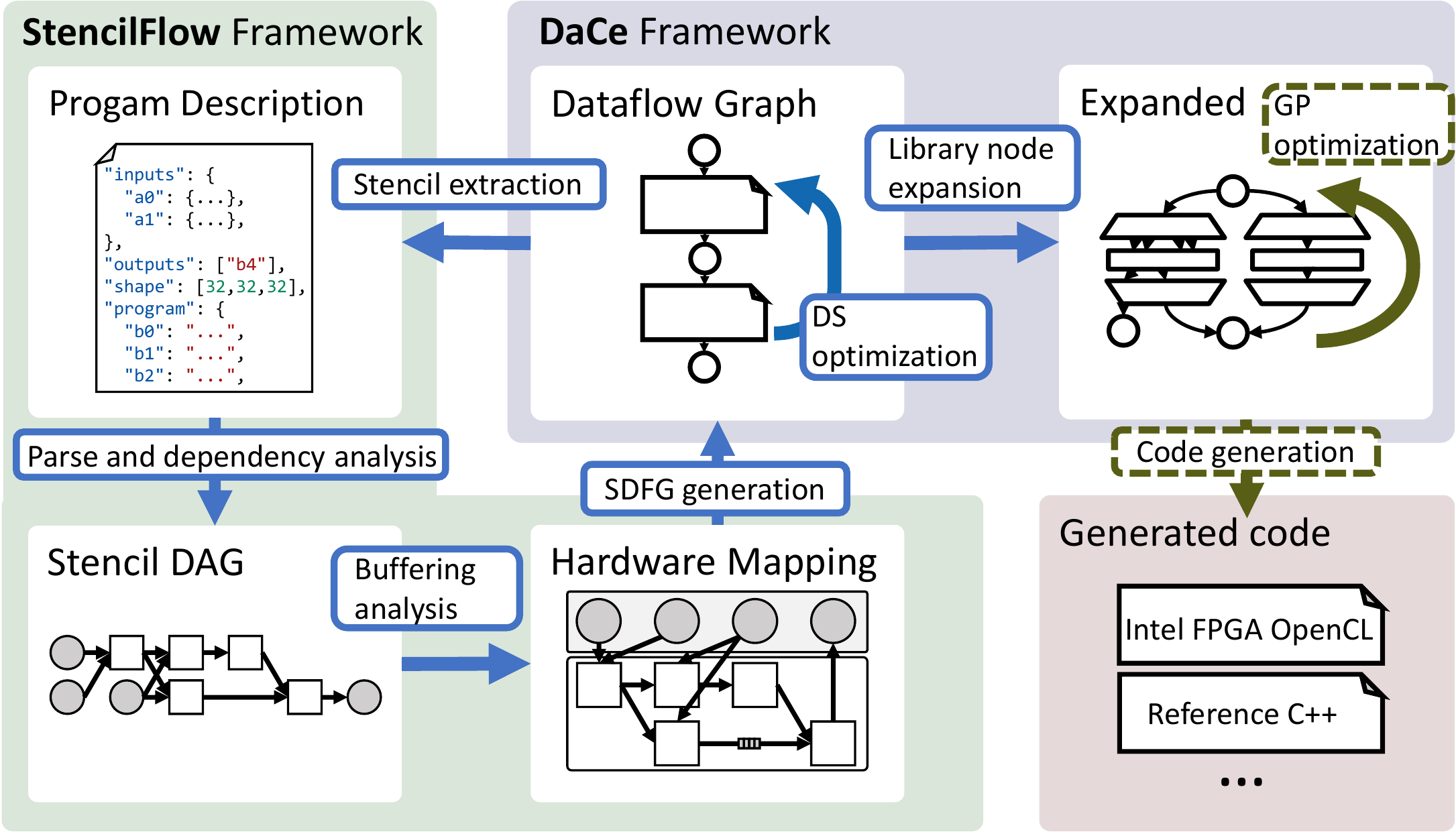}
    \caption{Workflow overview, with code artifacts annotated on arrows. Dashed outline indicates an existing feature that was extended.}
    \label{fig:workflow}
    \vspace{-1em}
\end{figure}

To summarize the stack described throughout the above, an overview of the StencilFlow workflow is shown in \figureref{workflow}. The StencilFlow framework is a pure Python code (${\approx}\SI{5300}{}$ SLOC at the time of writing) developed for the purpose of this work. The DaCe framework was extended with Python and C++ features to support domain-specific nodes.

The input program to StencilFlow can either be given as the JSON-based program description described in \secref{scope}, or as a DaCe dataflow graph containing domain-specific stencil nodes. In the latter case, we developed software that performs canonicalization passes to the DaCe graph, before extracting the stencil pattern to the standard program description format. This allows us to read in external programs, which will be required for the case study in \secref{horizontal_diffusion}.

StencilFlow can directly run the stencil program from the input description, transparently executing parsing, dependency analysis, buffering analysis, SDFG generation, domain-specific optimization, library node expansion, general purpose optimization, code generation, compilation of the host code, compilation of the kernel (requiring the full synthesis, placement and routing flow if FPGAs are targeted), execution of the program, and validation of results.

\section{Benchmarks}
\label{sec:experiments}

We benchmark the architectures emitted by StencilFlow to establish the highest achievable performance and bandwidth on a testbed platform, which we can use to analyze the characteristics required to push performance of stencil applications.

\subsection{Computing Expected Runtime}
\label{sec:performance_model}
\label{sec:expected_runtime}

We annotate benchmarks with the ``expected'' runtime, given by the lower bound on number of cycles required to evaluate the program, assuming all data is available at the earliest possible cycle. Because the full stencil DAG is executed in a pipeline parallel manner, we can model the runtime as a single, global pipeline. It is generally true for a pipelined circuit that the number of cycles required to process $N$ inputs is
\begin{align}
    C = L + I \cdot N\text{, }
    \label{eq:runtime}
\end{align}
where $L$ is the latency of the pipeline, and $I$ is the initiation interval (i.e., the number of cycles between allowing a new set of inputs to the circuit)~\cite{hls_transformations}. All architectures emitted by StencilFlow are fully pipelined, so we fix $I{=}\SI[per-mode=fraction]{1}{\cycles\per\operand}$. $N$ is the product of the domain dimensions (number of iterations in the iteration space), divided by the vectorization width $W$ when applicable.
$L$ is computed from the circuit latency and initialization delay described in \secref{delay_buffers}. $N$ and $L$ compose differently: $N$ covers the streaming section where stencils can operate in a pipeline parallel fashion, whereas $L$ covers the initialization phase where stencil units are not feeding downstream consumers. The depth of the DAG thus adversely affects the performance upper bound, while the size of the domain affects it favorably, increasing the relatively number of ``useful'' cycles to cycles spent in initialization. Since $L$ is only proportional to $D{-}1$ or fewer dimensions (see \secref{delay_buffers}), it becomes negligible when the domain is large relative to the depth of the stencil DAG. However, we include it when computing expected runtime for completeness.

\subsection{Experimental Platform}
\label{sec:experimental_setup}

To evaluate the efficiency of dataflow architectures laid out by StencilFlow, we map them to a state-of-the-art FPGA platform. We target the BittWare 520N PCI-e attached board, with an Intel GX~2800 Stratix~10 processor, 4 DDR4 memory banks with a combined peak bandwidth of $\SI{76.8}{\giga\byte\per\second}$, and four network-attached QSFP ports rated at $\SI{40}{\giga\bit\per\second}$. The annotated OpenCL code generated from DaCe is compiled with version 19.1.0 of the Intel FPGA OpenCL SDK and Quartus compiler, targeting the \texttt{p520\_max\_sg280l} shell offered by BittWare. This shell supports networking via OpenCL channels, which we target using the SMI library (\secref{smi}). The FPGAs are installed in the Noctua cluster at the Paderborn Center for Parallel Computing, which exposes a programmable, fully connected optical switch, allowing us to chain FPGAs together in a sequence with two $\SI{40}{\giga\bit\per\second}$ links between each consecutive device to explore multi-device scaling. Our benchmarks focus on 32-bit precision, as this is used in production by our motivating weather simulation example, and because this precision is supported natively on the Stratix~10. However, all parts of the StencilFlow stack support any data type recognized by the underlying compiler, including double precision floating point and integer types.

\subsection{Iterative Stencil Performance}
\label{sec:performance}
\label{sec:performance_benchmark}

\begin{figure}[b]
    \begin{minipage}{\columnwidth}
        \centering
        \includegraphics[width=\columnwidth]{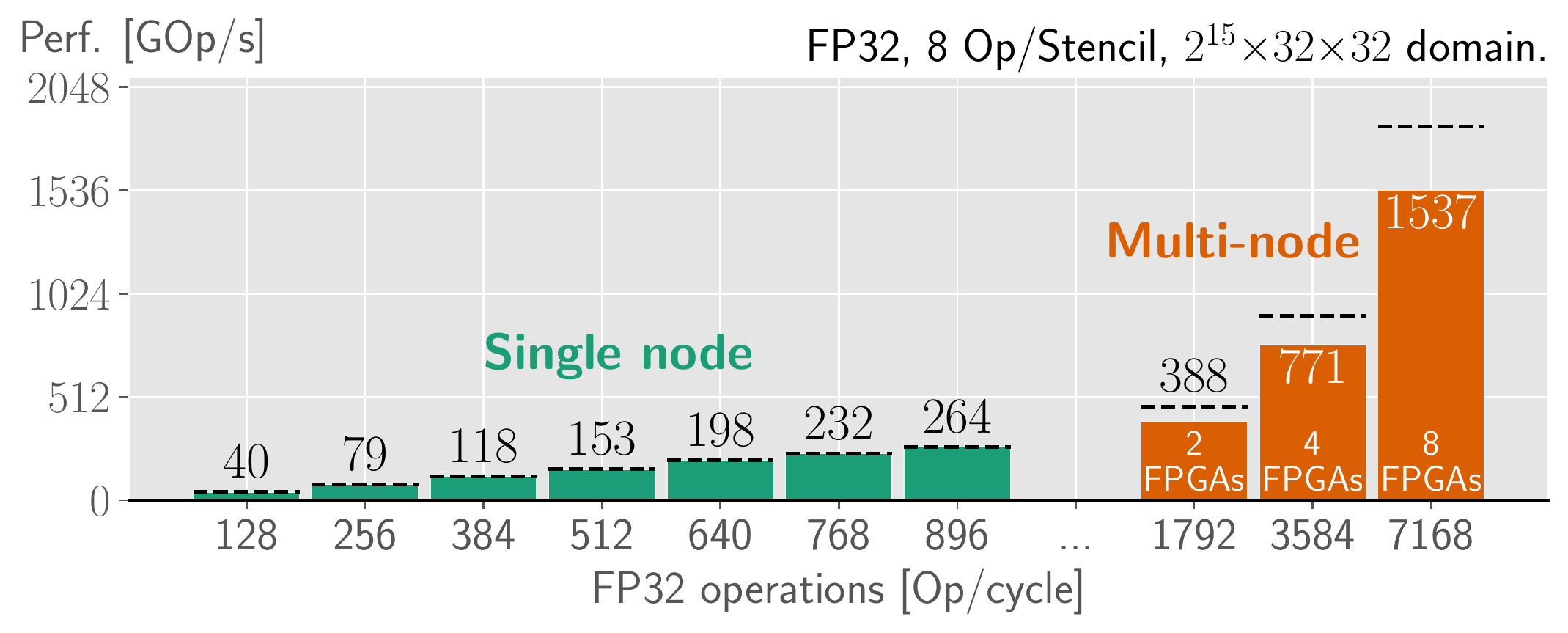}
        \caption{Performance scaling for single and multi-node.}
        \label{fig:performance}
    \end{minipage}\\
    \begin{minipage}{\columnwidth}
        \includegraphics[width=\columnwidth]{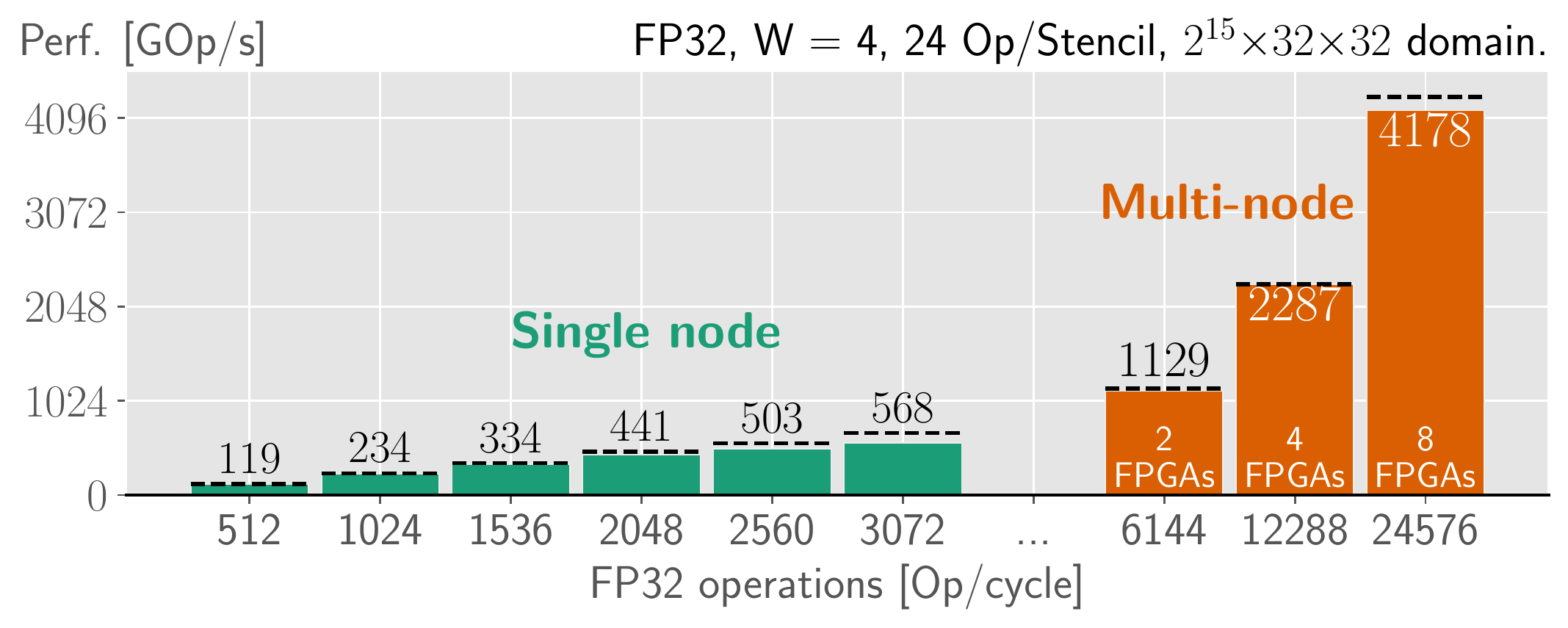}
        \caption{Performance scaling with 4-way vectorization.}
        \label{fig:performance_vectorized}
    \end{minipage}
\end{figure}

StencilFlow is built to handle complex stencil kernels, but is also capable of processing traditional, iterative-style stencil codes. We produce benchmarks using such kernels to establish the highest floating point performance reachable by StencilFlow, which can be compared to previous work. This is achieved by chaining together long linear sequences of stencils executed on a large input domain, analogous to time-tiled iterative stencils.
To evaluate scaling behavior of an iterative stencil, we gradually increase the
number of chained stencil computations until a single FPGA device is fully
utilized, then we continue the chain across multiple devices by replacing
accesses to on-chip FIFOs with network channels. We repeat the experiment with
and without vectorization, to see the effect of coarsening stencil stages. The
resulting benchmarks are shown in \figureref{performance} and
\figureref{performance_vectorized} without and with vectorization, respectively.

\begin{table}[b]
  \footnotesize
  \centering
  \resizebox{\columnwidth}{!}{\begin{tabular}{crrrrr}
  \toprule
  & \textbf{Performance} & \textbf{ALM} &\textbf{FF}  & \textbf{M20K} & \textbf{DSP} \\
  \midrule
    Total & & $103\:\text{M}$ &$3.7\:\text{M}$ & $11.7\:\text{K}$ &$5760$ \\
    Avail. & & $692\:\text{K}$ &$2.8\:\text{M}$ & $8.9\:\text{K}$ &$4468$ \\
  \midrule
  
       Jacobi 3D &  \multirow{2}{*}{$\SI{265}{\giga\op\per\second}$} & $233\:\text{K}$ &	$534\:\text{K}$ &	1495 &	784 \\
    (Ours) & & 33.6\% &	19.3\% &	16.7\% &	17.6\% \\[.5em]
    
    Jacobi 3D & \multirow{2}{*}{$\SI{921}{\giga\op\per\second}$} & $437\:\text{K}$ &	$1207\:\text{K}$ &	2285 &	3072 \\
    $W{=}8$ (Ours) & & 63.1\% &	43.6\% & 25.5\% & 68.8\% \\[.5em]
  
    Diffusion 2D & \multirow{2}{*}{$\SI{1313}{\giga\op\per\second}$} & $449\:\text{K}$ &	$1329\:\text{K}$ &	$2565$ & $2304$ \\
    $W{=}8$ (Ours) & & $64.8\%$ &	$48.0\%$ &	$28.6\%$ &	$51.6\%$ \\[.5em]
    
   Diffusion 3D & \multirow{2}{*}{$\SI{1152}{\giga\op\per\second}$} & $567\:\text{K}$ &	$1606\:\text{K}$ &	$5357$ & $3072$ \\
    $W{=}8$ (Ours) & & $81.9\%$ &	$57.9\%$ &	$59.8\%$ &	$68.8\%$ \\[.5em]
    
    \midrule
    Diffusion 2D & \multirow{2}{*}{$\SI{913}{\giga\op\per\second}$} & $471.4\:\text{K}$ &	$1173.6\:\text{K}$ &	2204 &	3844 \\
    (Zohouri~et. al.~\cite{zohouri_stencil}) & & 68.0\% & 42.3\% & 24.6\% & 86.0\% \\
    Diffusion 3D & \multirow{2}{*}{$\SI{934}{\giga\op\per\second}$} & $450.5\:\text{K}$ &	$1078.2\:\text{K}$ &	8684 &	3592 \\
    (Zohouri~et. al.~\cite{zohouri_stencil}) & & 65.0\% & 38.9\% & 97.0\% & 80.4\% \\
    \midrule
    Waidyasooriya & \multirow{2}{*}{$\SI{630}{\giga\op\per\second}$} & \multicolumn{4}{c}{\multirow{2}{*}{Arria 10 GX 1150}} \\
    and Hariyama~\cite{waidyasooriya2019multi} & & & & & \\[0.5em]
    SODA~\cite{soda} & $\SI{135}{\giga\op\per\second}$ & \multicolumn{4}{c}{ADM-PCIE-KU3} \\[0.5em]
    Niu et al.~\cite{runtime_stencil_trets} & $\SI{119}{\giga\op\per\second}$ &
    \multicolumn{4}{c}{Virtex-6 SX475T} \\[0.5em]
    Ben-Nun et al.~\cite{dapp} & $\SI{139}{\giga\op\per\second}$ &
    \multicolumn{4}{c}{Virtex UltraScale+ VCU1525} \\
  \bottomrule
  \end{tabular}}
  \caption{Highest performing kernels and their resource usage.}
  \label{tab:peak_measured}
  \vspace{1em}
\end{table}

\emph{Without vectorization}, the highest performing bitstream yields $\SI{264}{\giga\op\per\second}$ on a single device, and scales up to $\SI{1.5}{\tera\op\per\second}$ across 8 FPGAs. A 4-way vectorized code reaches $\SI{568.2}{\giga\op\per\second}$ and $\SI{4.2}{\tera\op\per\second}$ on single and multi-device, respectively.
Vectorization thus proves to be crucial to achieve high utilization of compute resources on the Stratix~10, as it reduces the ratio of overhead logic to computational logic. This further motivates the necessity of the stencil fusion transformation (\secref{stencil_fusion}) on input programs to coarsen the granularity of stencil nodes. Frequencies across all benchmarks are consistently in the range $292$-$\SI{317}{\mega\hertz}$, which is factored into the upper bound calculation shown as dashed black lines, computed from \equref{runtime} as $C/f$, where $f$ is the design frequency.

We additionally measure the highest performance achievable \emph{without}
networking on a single device, as we are unable to vectorize the stencils in the
distributed experiment further due to the network bandwidth bottlenecking the
computation, included in \tableref{peak_measured}.  As a non-relative measure of
device utilization, the table includes resource usage for the maximum performing
stencil for each data type. The highest measured stencil performance of
$\SI{1.3}{\tera\op\per\second}$ and $\SI{4.2}{\tera\op\per\second}$ marks a
$9.4{\times}$ and $30{\times}$ speedup over the stencil performance reported for
a single VCU1525 device in the original work on DaCe for single-device and
multi-device, respectively (which in turn outperformed a state-of-the-art HLS
compiler by five orders of magnitude, showing in inability of HLS compilers to
yield satisfactory out-of-the-box performance). 

For a more direct comparison on the Stratix 10 platform, we compare StencilFlow
to a handwritten stencil implementation. Zohouri~et~al.~\cite{zohouri_stencil}
combine spatial and temporal blocking in an HLS-based design to achieve high
performance on stencil codes on an Arria~10 FPGA. We extend the authors' work by
building their code\footnote{\url{https://github.com/zohourih/Diffusion_FPGA},
commit 96588e2.} for Stratix~10, using the Diffusion 2D and 3D stencil codes. On
advice from the authors, we configure parameters to a vectorization width of
$16$, run enough repetitions that the kernel runs for multiple seconds to hide
initialization overhead, and disable burst interleaving. We include the
resulting performance in \tableref{peak_measured}, along with other previous
results by Niu~et~al.~\cite{runtime_stencil_trets} and Waidyasooriya and
Hariyama~\cite{waidyasooriya2019multi}, showing that StencilFlow is competitive
even with hand-tuned code. We also consider frameworks emitting stencil FPGA
code, including the Jacobi 3D result of SODA~\cite{soda}, which is the stencil
backend of HeteroHalide~\cite{heterohalide} and HeteroCL~\cite{heterocl}. For
previous work we note the FPGA used for evaluation by the respective authors. We
do not compare quantitatively to HeteroCL and
Wang~and~Liang~\cite{opencl_framework}, as the authors do not report absolute
performance numbers.

\subsection{Off-Chip Memory Bandwidth}
\label{sec:memory_bandwidth}
\label{sec:memory_benchmark}

\begin{figure}[b]
    \centering
    \includegraphics[width=\columnwidth]{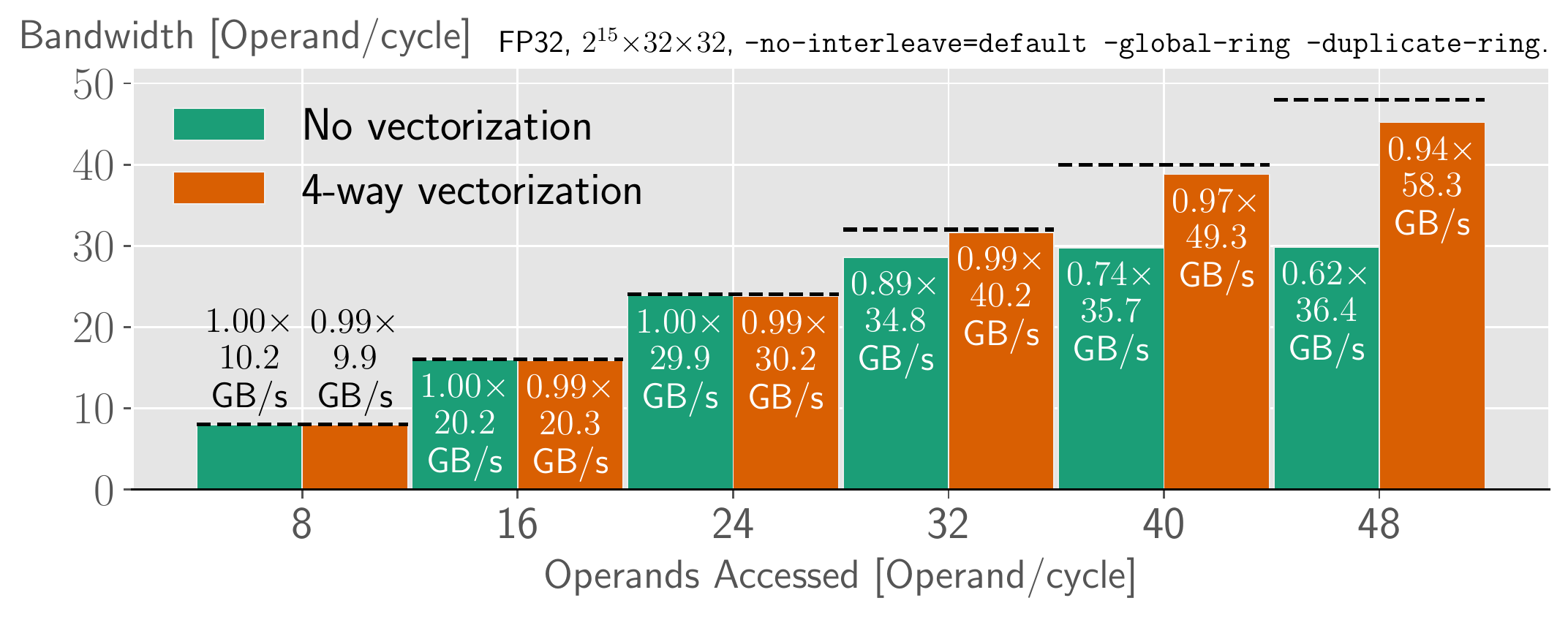}
    \caption{Effective bandwidth with number of operands requested per cycle
    (i.e., number of operands served if infinite bandwidth).}
    \label{fig:bandwidth}
    \label{fig:memory_benchmark}
\end{figure}

To measure achievable off-chip memory bandwidth by StencilFlow programs, we run two series of benchmarks: first, we measure the effective bandwidth utilization when scaling up \emph{number of accesses}, but accessing only 32-bits per cycle at each access point. This stresses the routing on the device to deliver data to all end-points every cycle. Second, we request the same total number of 32-bit operands, but at fewer, vectorized endpoints, requiring more operands per cycle per endpoint. We found the \texttt{-global-ring} and \texttt{-duplicate-ring} options to the Intel FPGA OpenCL compiler to significantly increase the number of parallel access points supported in the architecture before designs dropped in frequency. The resulting benchmarks, along with the analytically computed performance upper bound, are shown in \figureref{bandwidth}. For the non-vectorized green bars, the x-axis corresponds to the number of access points, while the number of access points for the vectorized orange bars is the number of operands divided by the vector size of 4 (i.e., up to 12 access points are depicted).

After 24 parallel access points, we see a decrease in effective memory performance relative to peak, flattening out at $\SI{36.4}{\giga\byte\per\second}$, which is $36.4/76.8=47\%$ of peak bandwidth. This marks the limit of the memory controller crossbar, and of routing a large number of memory accesses across the device. The 4-way vectorized scenario allows for higher achievable bandwidth, but experiences a drop in efficiency at a lower number of access points ($0.94{\times}$ at 12 access points), and flattens out at $\SI{58.3}{\giga\byte\per\second}$, which is $76\%$ of peak bandwidth. No further increase was seen with more access points, and 8-way vectorized programs achieve similar bandwidth.

\begin{figure*}
    \begin{minipage}{.5\linewidth}
    \begin{subfigure}{\linewidth}
        \centering
        \includegraphics[width=\linewidth]{figures/horizontal_diffusion_before.pdf}
        \caption{\small Input program as SDFG.}
        \label{fig:horizontal_diffusion_before}
    \end{subfigure}
    \\
    \begin{subfigure}{\linewidth}
        \centering
        \includegraphics[width=\linewidth]{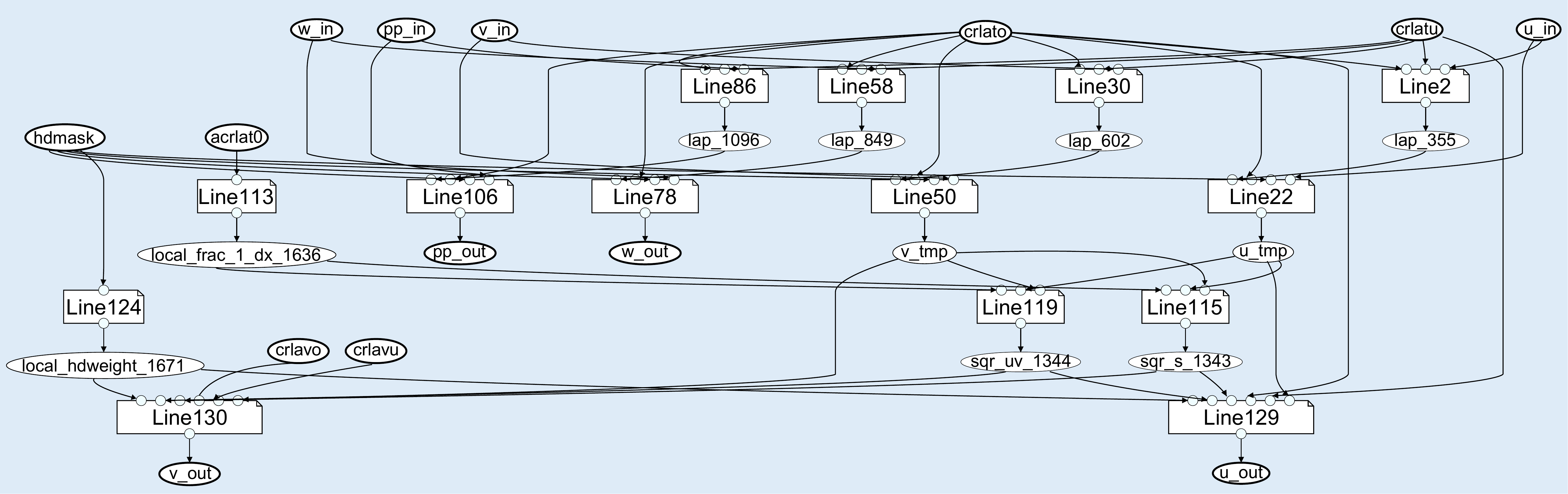}
        \caption{\small Canonicalized SDFG.}
        \label{fig:horizontal_diffusion_canon}
    \end{subfigure}
    \end{minipage}
    \qquad
    \begin{subfigure}{.5\linewidth}
        \centering
        \includegraphics[width=.73\linewidth]{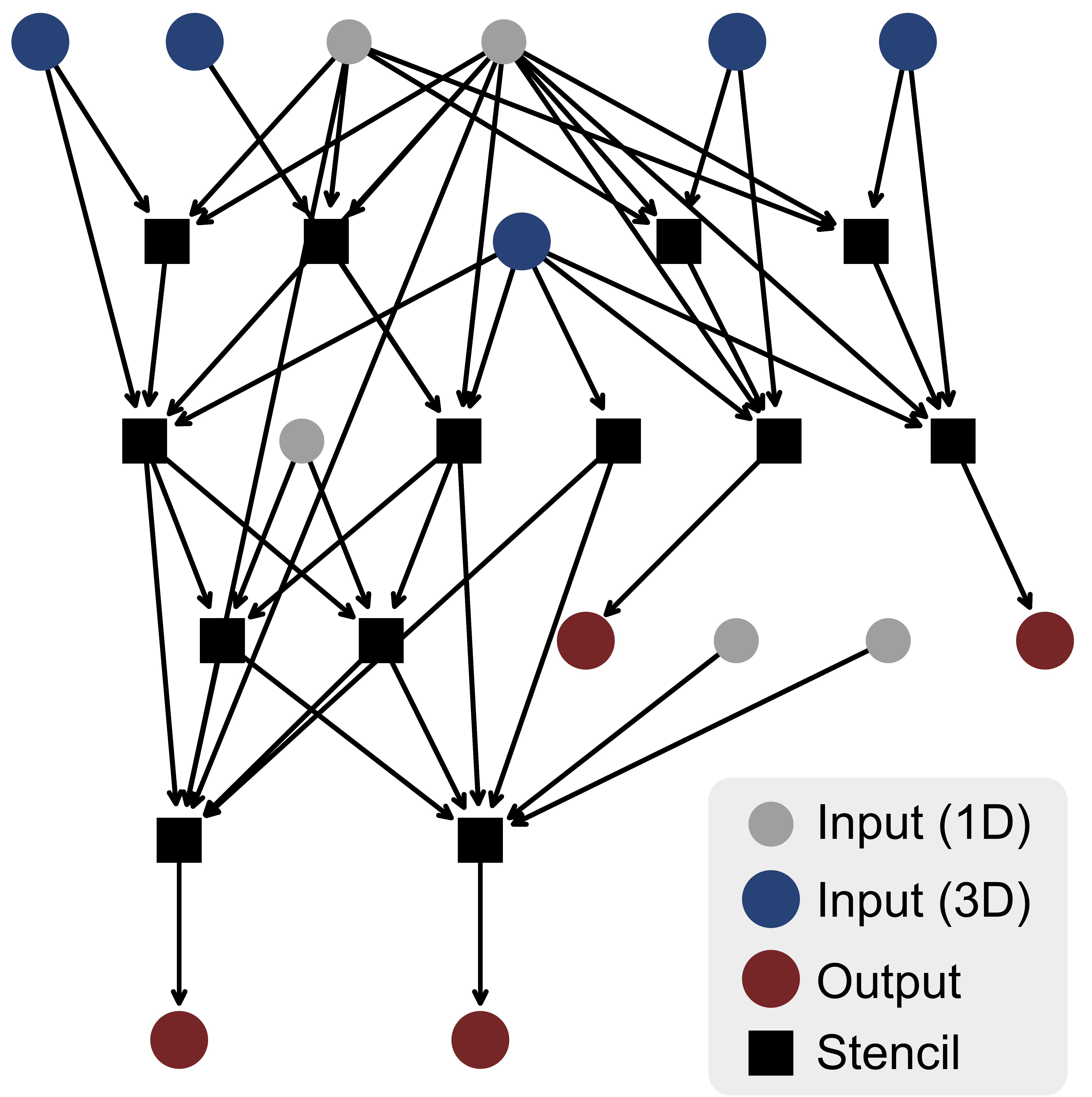}
        \caption{\small DAG of the transformed program inferred by StencilFlow.}
        \label{fig:horizontal_diffusion_df}
    \end{subfigure}
    \caption{Horizontal diffusion stencil program from the COSMO weather and climate model.}
    \label{fig:horizontal_diffusion}
    \vspace{-1em}
\end{figure*}

\section{Weather Simulation Application Study}
\label{sec:horizontal_diffusion}

To stress the full capability of the StencilFlow stack we evaluate the \emph{horizontal diffusion} stencil program, a large real-life weather simulation kernel from the COSMO weather model.
Horizontal diffusion is a 4th order explicit method performed on a staggered latitude-longitude grid with Smagorinsky diffusion to smoothen wind velocity components~\cite{smagorinsky}.
We obtain the program from an input SDFGs using stencil library nodes, shown for horizontal diffusion in \figureref{horizontal_diffusion_before}, applying the \texttt{NestDim} and \texttt{MapFission} transformations described in \secref{dace_extensions}, resulting in an SDFG as the one shown in \figureref{horizontal_diffusion_canon}, from which the stencil program is extracted. The DAG in \figureref{horizontal_diffusion_df} is created after aggressively fusing consecutive stencils (see~\secref{stencil_fusion}).
In the fully fused program, initialization latency ($L$ in \equref{runtime}) accounts for ${\sim}0.7\%$ of the total number of iterations required to evaluate the program, and is thus a negligible overhead.
This program is run in production by the Swiss Federal Office of Meteorology and Climatology (MeteoSwiss), where simulations are performed with 32-bit floating point on an NVIDIA Pascal Tesla K80 cluster. We compare StencilFlow to the stronger TSMC~$\SI{16}{\nano\meter}$ Tesla P100 GPU on the same architecture (comparable release window to the Stratix~10), a TSMC~$\SI{12}{\nano\meter}$ Tesla V100 Volta GPU, and a 12-core Xeon CPU.

\vspace{-0.25em}
\subsection{Horizontal Diffusion Analysis}
\label{sec:program_analysis}

The horizontal diffusion DAG characterized by a high number of stencils reading the same input locations (28 accesses of 10 unique fields), allowing for the communication volume between them to be consolidated via delay buffers, as well as complex dependencies between stencil nodes (each non-source stencil receives data from $2{-}6$ other stencil nodes). This requires the full complexity of an arbitrary DAG, and allows us to stress the full stack of StencilFlow. 

Floating point operations in the DAG include 87 additions, 41 multiplications, and 2 square roots, in addition to 2 minimum and 2 maximum operations, and ternary operations resulting in 20 data-dependent branches. With maximum reuse of all input fields and all computed fields (i.e., perfect locality), the program reads $5 I J K + 5 I$ operands and writes $4 I J K$ operands, for a total of $9 I J K + 5 I$ operands. Considering floating point arithmetic only, this implies a upper bound arithmetic intensity of (square root is counted as one operation):
\vspace{-0.5em}\begin{align*}
    \frac{(87 + 41 + 2) I J K \; \left[\si{\ops}\right]}{9 I J K + 5 I \; \left[\si{\operands}\right]}
    \approx \frac{130}{9}\;\left[\si[per-mode=fraction]{\ops\per\operand}\right]\text{, }
\end{align*}
which for $32$-bit floating point corresponds to
\begin{align}
    \frac{130/9 \;\left[\si{\ops\per\operand}\right]}{4 \; \left[\si{\byte\per\operand}\right]} = \frac{65}{18} \; \left[\si[per-mode=fraction]{\ops\per\byte}\right]\text{.}
    \label{eq:arithmetic_intensity}
\end{align}
Using the benchmark of practically achievable bandwidth presented in \secref{memory_benchmark} for the Stratix~10 FPGA, the highest achievable performance in roofline model~\cite{roofline} terms is:
\begin{align}
    \frac{65}{18} \; \left[\si{\ops\per\byte}\right] \cdot 58.3 \; \left[\si{\giga\byte\per\second}\right] = 210.5 \; \left[\si{\giga\op\per\second}\right]\text{,}
    \label{eq:horizontal_diffusion_max_performance}
\end{align}
or $\SI{277.3}{\giga\op\per\second}$ at the peak data sheet bandwidth of
$\SI{76.8}{\giga\byte\per\second}$. This is well below what is achievable by a
stencil program with higher arithmetic intensity (see
\secref{performance_benchmark}), indicating that high bandwidth is required to
shine in realistic stencil applications. We compute the bandwidth required to
saturate the compute performance measured in \secref{performance_benchmark} for
the arithmetic intensity of the studied program to be:
\begin{align}
    \frac{917.1\;\left[\si{\giga\op\per\second}\right]}{65/18\;\left[\si{\op\per\byte}\right]} = 254.0 \; \left[\si{\giga\byte\per\second}\right]\text{.}
    \label{eq:horizontal_diffusion_bandwidth_required}
\end{align}
The ideal logic to bandwidth ratio is thus off from the ideal ratio by a factor of ${\sim}3{-}4$ on the target Stratix~10 platform.
To explore the performance potential of the Stratix~10 without this memory bottleneck, we will include experiments with simulated ``infinite'' memory bandwidth, by replacing memory accesses with compile-time constants fed to the computational circuit (and omitting validation of functional correctness). 

\subsection{Horizontal Diffusion Benchmark}
\label{sec:results}

We compile the DAG in \figureref{horizontal_diffusion_df} for the Stratix~10
from the constructed dataflow graph by StencilFlow. As shown in the analysis
above, the program is bandwidth-bound on this platform, which requires us to
saturate the bandwidth to maximize performance. Without vectorization, the
pipelined circuit requires approximately $\SI{9}{\operands\per\cycle}$,
corresponding to $\SI{10.8}{\giga\byte\per\second}$ at $\SI{300}{\mega\hertz}$
for single precision floating point. We thus vectorize the program by a factor
of 8 for a maximum bandwidth of $\SI{86.4}{\giga\op\per\second}$, in addition to
building a 16-way vectorized kernel with simulated input memory to evaluate
performance without the memory bottleneck.  We target a
$128{\times}128{\times}80$ domain size, which is used for performance
benchmarking by MeteoSwiss. Specifically, a $128{\times}128$ horizontal domain
is stacked in $80$ vertical layers.  In addition to runtime and the effective
performance, we consider peak memory bandwidth and the associated fraction of
highest achievable performance \emph{for the given arithmetic intensity}
computed according to \equref{arithmetic_intensity} (\%Roof.). The results are
listed in \tableref{horizontal_diffusion_results}.

\begin{table}[h]
    \centering
    \resizebox{\columnwidth}{!}{\begin{tabular}{l | r r r r r r}
        & Runtime & Performance & Peak BW. & \%Roof. \\\hline
        Stratix~10 & $\SI{1178}{\micro\second}$ & $\SI{145}{\giga\op\per\second}$ & $\SI{77}{\giga\byte\per\second}$ & $52\%$ \\
        Stratix~10$^*$ & $\SI{332}{\micro\second}$ &
        $\SI{513}{\giga\op\per\second}$ & $\infty\:\si{\giga\byte\per\second}$ &
        $-\:\;\;$ \\
        Xeon 12C & $\SI{5270}{\micro\second}$ & $\SI{32}{\giga\op\per\second}$ & $\SI{68}{\giga\byte\per\second}$ & $13\%$ \\
        P100 & $\SI{810}{\micro\second}$ & $\SI{210}{\giga\op\per\second}$ & $\SI{732}{\giga\byte\per\second}$ & $8\%$ \\
        V100 & $\SI{201}{\micro\second}$ & $\SI{849}{\giga\op\per\second}$ & $\SI{900}{\giga\byte\per\second}$ & $26\%$ \\
    \end{tabular}}
    \centering
    \\[0.5em]{\small$^*$Without memory bandwidth constraints.}
    \caption{Horizontal diffusion benchmarks.}
    \label{tab:horizontal_diffusion_results}
    \vspace{-0.5em}
\end{table}

We include CPU and GPU performance as a point of comparison, using a 12-core
Intel Xeon $2.60$/$\SI{3.50}{\giga\hertz}$ E5-2690V3 CPU, and NVIDIA Tesla P100
and V100 GPUs, compiled with CUDA v10.1 and gcc 8.3.0. The application is
synthesized using the MeteoSwiss Dawn~\cite{dawn} stencil-optimizing compiler
toolchain\footnote{\url{https://github.com/MeteoSwiss-APN/dawn}, commit
\texttt{4ae6dc0}.}, which was also used to generate the StencilFlow input
program. Dawn is specifically designed to optimize weather and climate stencil
programs for GPU and CPU, employing data movement optimizations, GPU kernel
fusion, CPU multi-threading, vectorization, and efficient GPU boundary
scheduling. The domain size of $128{\times}128{\times}80$ is sufficient for
saturating the GPU thread scheduler (i.e., larger domains do not significantly
increase GPU performance). The horizontal diffusion program emitted by Dawn for
CPU and GPU executes five components of horizontal diffusion as distinct
kernels. We omit kernel launch overhead and report the raw kernel execution time
only, included in \tableref{horizontal_diffusion_results}. 

The FPGA platform outperforms the CPU by $4.5{\times}$ and is outperformed by either GPU, but comes closest to the upper bound (\equref{horizontal_diffusion_max_performance}) imposed by its roofline characteristics at the given arithmetic intensity: $52\%$ of the bandwidth upper bound ($69\%$ of the highest measured bandwidth in \secref{memory_benchmark}), at $26\%$ ALMs, $27\%$ DSPs, and $20\%$ M20K utilization, respectively. The benchmark simulating infinite memory bandwidth shows significant headroom for pushing the performance at this arithmetic intensity with higher bandwidth off-chip memory: without the bandwidth bottleneck, the Stratix~10 would outperform the P100, but falls at $60\%$ of the performance of the V100, at $46\%$ ALMs, $48\%$ DSPs, and $20\%$ M20Ks. 

\subsection{Silicon Efficiency}

The Stratix~10 is estimated to be a $\SI{700}{\milli\meter^2}$ die~\cite{stratix10_size} (half the Stratix~10M, which fuses two Stratix~10 chiplets) on Intel's $\SI{14}{\nano\meter}$ process, compared to $\SI{610}{\milli\meter^2}$ on TSMC~$\SI{16}{\nano\meter}$ and $\SI{815}{\milli\meter^2}$ on TSMC~$\SI{12}{\nano\meter}$ for the P100 and V100, respectively. Using the benchmarks from \tableref{horizontal_diffusion_results}, this amounts to a silicon efficiency of $\SI{0.21}{}$ and $\SI{0.71}{}$ $\frac{\si{\giga\op\per\second}}{\si{\milli\meter\squared}}$ with and without the memory bottleneck for the Stratix~10, respectively; $\SI{0.34}{}\frac{\si{\giga\op\per\second}}{\si{\milli\meter\squared}}$ for P100; and $\SI{1.04}{}\frac{\si{\giga\op\per\second}}{\si{\milli\meter\squared}}$ for the V100, when performing the horizontal diffusion experiment.

\subsection{Spatial Tiling}
\label{sec:fast_memory}

We have not considered spatial tiling, as on-chip memory requirements were not a restriction for building the large weather stencil program evaluated. Both memory bandwidth and logic were bottlenecks before on-chip memory capacity, despite minimizing off-chip memory bandwidth in the program. Eventually, increasing the domain size will scale the internal buffer and delay buffer sizes beyond what is feasible to buffer in on-chip memory. Spatial tiling can be employed in this scenario, introducing redundant computation at the domain boundaries proportional to the DAG depth and the tile surface-to-volume ratio. This is primarily a scheduling challenge, which can be efficiently solved in practice~\cite{zohouri_stencil}.

\section{Related Work}
\label{sec:related_work}

There are numerous works on stencil accelerators on FPGAs~\cite{fu_clapp, zohouri_stencil, waidyasooriya2019multi}, including for multi-device settings on up to 9 interconnected FPGAs~\cite{sano_multi_fpga}, all of which we have considered throughout this work. Other frameworks generating stencil architectures have also been proposed~\cite{soda, heterocl, heterohalide, opencl_framework}, which we consider in \secref{performance_benchmark}. Common to these works is that they treat a single stencil operation applied iteratively, allowing them to unroll the time dimension as a source of temporal locality. StencilFlow is on a par or outperforms all the above on simple iterative stencils, and treats a much wider range of input programs.
Niu~et~al.~\cite{runtime_stencil_trets} explore runtime reconfiguration of an FPGA to eliminate idle operators during program execution. Runtime reconfiguration is not beneficial for stencil programs considered by StencilFlow, as all operators are assumed to operate in the same iteration space and fully in parallel after the initialization phase.

Darkroom~\cite{darkroom} is a framework producing spatial accelerators of image processing pipelines from a high-level input DSL. StencilFlow takes a similar approach, but accepts a wider scope of input programs: in particular arbitrary DAGs of stencils, and 3D input/output domains. Other DSLs~\cite{stella, lift_stencil} do not consider spatial computing architectures.

For the application study, Singha~et~al.~\cite{narmada, nero} present a hand-tuned implementation of the horizontal diffusion application targeting an FPGA+CPU coherent system. The authors report $\SI{129.9}{\giga\op\per\second}$ on an ADM-PCIE-9V3 board with the NARMADA accelerator, and $\SI{485.4}{\giga\op\per\second}$ on an ADM-PCIE-9H7 board with the NERO accelerator, the latter owing its large increase in performance to the introduction of HBM memory, effectively eliminating the memory bottleneck described by \equref{horizontal_diffusion_bandwidth_required}. The fully code generated kernels emitted by StencilFlow outperform the DDR4-based accelerator when memory bound, and the HBM-based when compute bound (i.e., when high memory bandwidth is simulated).

\section{Conclusion}
\label{sec:conclusion}

We introduced StencilFlow, an end-to-end analysis, optimization and code-generation stack built on the DaCe framework, enabling the generation of complex high-performance stencil programs on spatial architectures from a high-level input DSL.
Based on a DAG representation, StencilFlow automatically insert buffers within and between stencil operations to achieve \emph{perfect reuse} of all data in the program.
Architectures emitted by StencilFlow achieve the \emph{highest recorded single-device performance} of $\SI{1.31}{\tera\op\per\second}$, and the \emph{highest recorded multi-device performance} of $\SI{4.18}{\tera\op\per\second}$ on 8 FPGAs.
We demonstrated the domain complexity supported by the framework by treating a large stencil program used in production for weather prediction, comparing the generated architecture to state-of-the-art GPU and CPU performance.
We release StencilFlow as open source software, enabling reproducibility and allowing scientists to easily target spatial computing accelerators with complex stencil programs.

\section*{Acknowledgements}

\noindent We authors wish to thank Tobias Kenter, Christian Plessl, and the
Paderborn Center for Parallel Computing (PC$^2$) for generously providing
support and compute hours on the Noctua FPGA cluster; the Swiss National
Supercomputing Center (CSCS) for providing computing infrastructure; and Jakub
Ber\'{a}nek for last-minute engineering contributions.  This work was supported
by the European Research Council under the European Union's Horizon 2020
programme (grant agreement DAPP, No. 678880). Tal Ben-Nun is supported by the
Swiss National Science Foundation (Ambizione Project No. 185778).


\bibliographystyle{IEEEtran}
\bibliography{StencilFlow}

\end{document}